\documentclass[12pt]{article}
\usepackage{amsmath}

\usepackage{amssymb}
\usepackage{bm}
\usepackage{braket}
\usepackage[dvips]{graphicx}

\begin{document}

\baselineskip=17pt

\begin{titlepage}
\rightline{\tt arXiv:2203.05366}
\begin{center}
\vskip 1.5cm
\baselineskip=22pt
{\Large \bf {Correlation functions of scalar field theories}}\\
{\Large \bf {from homotopy algebras}}
\end{center}
\begin{center}
\vskip 1.0cm
{\large Yuji Okawa}
\vskip 1.0cm
{\it {Graduate School of Arts and Sciences, The University of Tokyo}}\\
{\it {3-8-1 Komaba, Meguro-ku, Tokyo 153-8902, Japan}}\\
okawa@g.ecc.u-tokyo.ac.jp
\vskip 2.0cm

{\bf Abstract}
\end{center}

\noindent
We present expressions for correlation functions of scalar field theories
in perturbation theory using quantum $A_\infty$ algebras.
Our expressions are highly explicit and can be used for theories both in Euclidean space
and in Minkowski space including quantum mechanics.
Correlation functions at a given order of perturbation theory can be calculated algebraically
without using canonical quantization or the path integral,
and we demonstrate it explicitly for $\varphi^3$ theory.
We show that the Schwinger-Dyson equations are satisfied
as an immediate consequence of the form
of the expressions based on quantum $A_\infty$ algebras.

\end{titlepage}

\tableofcontents

\section{Introduction}
\label{section-1}
\setcounter{equation}{0}

Homotopy algebras such as
$A_\infty$ algebras~\cite{Stasheff:I, Stasheff:II, Getzler-Jones, Markl, Penkava:1994mu, Gaberdiel:1997ia}
and $L_\infty$ algebras~\cite{Zwiebach:1992ie, Markl:1997bj}
have been playing a significant role in the construction of string field theory,
which can be seen most magnificently in the construction of closed string field theory
by Zwiebach~\cite{Zwiebach:1992ie}.
When we consider projections onto subspaces of the Hilbert space of the string,
homotopy algebras have turned out to provide useful tools.
The projection onto on-shell states
describes on-shell scattering amplitudes~\cite{Kajiura:2003ax},
the projection onto the physical sector leads
to mapping between covariant and light-cone string field theories~\cite{Erler:2020beb},
and the projection onto the massless sector is relevant
for the low-energy effective
action~\cite{Sen:2016qap, Erbin:2020eyc, Koyama:2020qfb, Arvanitakis:2020rrk, Arvanitakis:2021ecw}.

We can also describe quantum field theories using
homotopy algebras~\cite{Hohm:2017pnh, Jurco:2018sby, Nutzi:2018vkl, Arvanitakis:2019ald, Macrelli:2019afx, Jurco:2019yfd, Saemann:2020oyz}.
For scalar field theories the description in terms of homotopy algebras
is rather trivial, which reflects the fact that there are no gauge symmetries
in scalar field theories.
However, the relation between the action and on-shell scattering amplitudes
is universal, and the description of on-shell scattering amplitudes
in terms of homotopy algebras is nontrivial for scalar field theories
and provides new perspectives.

In quantum field theory, we also consider correlation functions.
Since there does not seem to be any immediate relation between correlation functions
and projections in homotopy algebras,
we may have an impression that homotopy algebras will not be useful
in describing correlation functions.
On the other hand, the description of on-shell scattering amplitudes
in terms of homotopy algebras
is based on the fact that Feynman diagrams are algebraically generated
in this approach~\cite{Kajiura:2003ax, Doubek:2017naz, Masuda:2020tfa},
and we expect that there is a way to generate Feynman diagrams for correlation functions
using homotopy algebras as well.
Furthermore, the Batalin-Vilkovisky formalism~\cite{Batalin:1981jr, Batalin:1983ggl, Schwarz:1992nx}
can be thought of as being dual to the homotopy algebra,
and correlation functions have been discussed
in the framework of the Batalin-Vilkovisky formalism~\cite{Gwilliam:2012jg, Chiaffrino:2021pob}.
Therefore, we again expect that there is a way to describe correlation functions
using homotopy algebras.
In this paper we demonstrate that it is indeed the case
that we can describe correlation functions in terms of homotopy algebras,
and we present highly explicit expressions for correlation functions of scalar field theories
in perturbation theory using quantum $A_\infty$ algebras.

The rest of the paper is organized as follows.
In section~\ref{section-2} we explain the description
of scalar field theories in terms of quantum $A_\infty$ algebras
and we present our formula for correlation functions.
In section~\ref{section-3} we calculate correlation functions
of the free theory and confirm that our formula reproduces Wick's theorem.
In section~\ref{section-4} we consider $\varphi^3$ theory
and we calculate correlation functions in perturbation theory explicitly.
We then show that the Schwinger-Dyson equations are satisfied
as an immediate consequence of the form
of the expressions based on quantum $A_\infty$ algebras
in section~\ref{section-5}.
In section~\ref{section-6} we consider scalar field theories in Minkowski space.
Section~\ref{section-7} is devoted to conclusions and discussion.

\section{Correlation functions from quantum $A_\infty$ algebras}
\label{section-2}
\setcounter{equation}{0}

We explain the description of scalar field theories in terms of $A_\infty$ algebras
in subsection~\ref{subsection-2.1}.
We then explain the coalgebra representation of $A_\infty$ algebras
in subsection~\ref{subsection-2.2},\footnote{
The coalgebra representation of $A_\infty$ algebras
is explained in detail, for example, in appendix~A of~\cite{Erler:2015uba}
and in~\cite{Koyama:2020qfb}.
We mostly follow the conventions used in these papers.
}
and we consider projections onto subspaces
in subsection~\ref{subsection-2.3}.
Finally in subsection~\ref{subsection-2.4}
we present our formula
for correlation functions in terms of quantum $A_\infty$ algebras.

\subsection{Scalar field theories in terms of $A_\infty$ algebras}
\label{subsection-2.1}

Let us first consider scalar field theories in Euclidean space.
Scalar field theories in Minkowski space will be discussed later in section~\ref{section-6}.
The action of the free theory is given by
\begin{equation}
\frac{1}{2} \int d^d x \, [ \, \partial_\mu \varphi (x) \,  \partial_\mu \varphi (x)
+m^2 \, \varphi (x)^2 \, ] \,,
\end{equation}
where $\varphi (x)$ is a real scalar field in $d$ dimensions
and $m$ is a real parameter.

To describe this action in terms of an $A_\infty$ algebra,
we introduce two copies of the vector space of functions of $x$.
We denote them by $\mathcal{H}_1$ and $\mathcal{H}_2$,
and we define $\mathcal{H}$ by
\begin{equation}
\mathcal{H} = \mathcal{H}_1 \oplus \mathcal{H}_2 \,.
\end{equation}
The vector space $\mathcal{H}$ is graded with respect to degree.
Any element in $\mathcal{H}_1$ is degree even
and any element in $\mathcal{H}_2$ is degree odd,
but signs from anticommuting degree-odd objects never appear
in the calculations to be presented in this paper.

We then introduce a symplectic form and denote it
by $\omega \, ( \, \varphi_1 (x), \varphi_2 (x) \, )$
for $\varphi_1 (x)$ and $\varphi_2 (x)$ in $\mathcal{H}$.
The symplectic form
for $\varphi_1 (x) $ in $\mathcal{H}_1$ and $\varphi_2 (x)$ in $\mathcal{H}_2$
is defined by
\begin{equation}
\omega \, ( \, \varphi_1 (x), \varphi_2 (x) \, ) = \int d^d x \, \varphi_1 (x) \, \varphi_2 (x) \quad
\text{for} \quad \varphi_1 (x) \in \mathcal{H}_1 \quad \text{and} \quad \varphi_2 (x) \in \mathcal{H}_2 \,.
\end{equation}
The symplectic form is graded antisymmetric,
and $\omega \, ( \, \varphi_1 (x), \varphi_2 (x) \, )$
for $\varphi_1 (x) $ in $\mathcal{H}_2$ and $\varphi_2 (x)$ in $\mathcal{H}_1$ is given by
\begin{equation}
\omega \, ( \, \varphi_1 (x), \varphi_2 (x) \, ) = {}-\int d^d x \, \varphi_1 (x) \, \varphi_2 (x) \quad
\text{for} \quad \varphi_1 (x) \in \mathcal{H}_2 \quad \text{and} \quad \varphi_2 (x) \in \mathcal{H}_1 \,.
\end{equation}
The symplectic form vanishes for other cases:
\begin{equation}
\omega \, ( \, \varphi_1 (x), \varphi_2 (x) \, ) = 0 \quad
\text{for} \quad \varphi_1 (x) \,, \varphi_2 (x)  \in \mathcal{H}_1 \quad
\text{or} \quad \varphi_1 (x) \,, \varphi_2 (x) \in \mathcal{H}_2 \,.
\end{equation}

The last ingredient for the action of the free theory is $Q$,
which is a linear operator on $\mathcal{H}$.
The action of $Q$ on $\varphi (x)$ in $\mathcal{H}_1$ is defined by
\begin{equation}
Q \, \varphi (x) = ( \, {}-\partial^2 +m^2 \, ) \, \varphi (x) \quad
\text{for} \quad \varphi (x) \in \mathcal{H}_1 \,,
\end{equation}
and $Q \, \varphi (x)$ is in $\mathcal{H}_2$.
On the other hand, the operator $Q$ annihilates any element in $\mathcal{H}_2$:
\begin{equation}
Q \, \varphi (x) = 0 \quad
\text{for} \quad \varphi (x) \in \mathcal{H}_2 \,.
\end{equation}
Let us summarize the nonvanishing part of $Q$ as follows:
\begin{equation}
Q: \mathcal{H}_1 \to \mathcal{H}_2 \,.
\end{equation}
Since an element in $\mathcal{H}_1$ is degree even
and an element in $\mathcal{H}_2$ is degree odd,
we say that $Q$ is degree odd.
Note that the operator $Q$ has the following cyclic property:
\begin{equation}
\omega \, ( \, \varphi_1 (x), Q \, \varphi_2 (x) \, )
= {}-(-1)^{{\rm deg} (\varphi_1)} \, \omega \, ( \, Q \, \varphi_1 (x), \varphi_2 (x) \, ) \,,
\label{Q-cyclicity}
\end{equation}
where
${\rm deg} (\varphi) = 0$ mod~$2$ when $\varphi (x)$ is degree even
and
${\rm deg} (\varphi) = 1$ mod~$2$ when $\varphi (x)$ is degree odd.
We also use this notation for operators and maps.
For example, we write ${\rm deg} (Q) = 1$ mod~$2$.
In the current case the symplectic form~\eqref{Q-cyclicity} can be nonvanishing only when
both $\varphi_1 (x)$ and $\varphi_2 (x)$ are in $\mathcal{H}_1$,
so the sign factor~$(-1)^{{\rm deg} (\varphi_1)}$ in~\eqref{Q-cyclicity} is trivial,
but it can be nontrivial
for more general $A_\infty$ algebras.
Using the symplectic form $\omega$ and the operator $Q$,
the action of the free theory can be written
for $\varphi (x)$ in $\mathcal{H}_1$ as follows:
\begin{equation}
\frac{1}{2} \int d^d x \, [ \, \partial_\mu \varphi (x) \,  \partial_\mu \varphi (x)
+m^2 \, \varphi (x)^2 \, ]
= \frac{1}{2} \, \omega \, ( \, \varphi (x), Q \, \varphi (x) \, ) \,. 
\end{equation}
Note that the operator $Q$ is nilpotent:
\begin{equation}
Q^2 = 0 \,.
\label{nilpotency}
\end{equation}
This relation is trivially satisfied in the current case
as the action of $Q$ can be nonvanishing only when it acts on an element in $\mathcal{H}_1$
but the resulting element is in $\mathcal{H}_2$ and is annihilated by the following action of $Q$.
For more general cases, this property of $Q$ is related
to the gauge invariance of the free theory.

Let us next consider interactions.
The classical action of $\varphi^3$ theory in Euclidean space $S^{(0)}$ is given by
\begin{equation}
S^{(0)} = \int d^d x \, \biggl[ \, \frac{1}{2} \, \partial_\mu \varphi (x) \, \partial_\mu \varphi (x)
+\frac{1}{2} \, m^2 \, \varphi (x)^2
-\frac{1}{6} \, g \, \varphi (x)^3 \, \biggr] \,,
\end{equation}
where $g$ is the coupling constant.
To describe the cubic interaction in terms of an $A_\infty$ algebra
we introduce a linear map from $\mathcal{H} \otimes \mathcal{H}$ to $\mathcal{H}$
and denote it by $b_2$.
The action of $b_2$ on $\varphi_1 (x) \otimes \varphi_2 (x)$
for $\varphi_1 (x)$ and $\varphi_2 (x)$ in $\mathcal{H}_1$ is defined by
\begin{equation}
b_2 \, ( \, \varphi_1 (x) \otimes \varphi_2 (x) \, )
= {}-\frac{g}{2} \, \varphi_1 (x) \, \varphi_2 (x) \quad
\text{for} \quad \varphi_1 (x) \in \mathcal{H}_1 \quad
\text{and} \quad \varphi_2 (x) \in \mathcal{H}_1 \,,
\label{b_2}
\end{equation}
and $b_2 \, ( \, \varphi_1 (x) \otimes \varphi_2 (x) \, )$ is in $\mathcal{H}_2$.
The action of $b_2$ for other cases vanishes:
\begin{equation}
b_2 \, ( \, \varphi_1 (x) \otimes \varphi_2 (x) \, )
= 0 \quad
\text{when} \quad \varphi_1 (x) \in \mathcal{H}_2 \quad
\text{or} \quad \varphi_2 (x) \in \mathcal{H}_2 \,.
\end{equation}
We can thus summarize the nonvanishing part of $b_2$ as follows:
\begin{equation}
b_2: \mathcal{H}_1 \otimes \mathcal{H}_1 \to \mathcal{H}_2 \,.
\end{equation}
It follows from this that $b_2$ is degree odd.
Note that the operator $b_2$ has the following cyclic property:
\begin{equation}
\omega \, ( \, \varphi_1 (x), b_2 \, ( \, \varphi_2 (x) \otimes \varphi_3 (x) \, ) \, )
= {}-(-1)^{{\rm deg} (\varphi_1)} \,
\omega \, ( \, b_2 \, ( \, \varphi_1 (x) \otimes \varphi_2 (x) \, ) \,, \varphi_3 (x) \, ) \,.
\label{cubic-cyclic}
\end{equation}
In the current case the symplectic form can be nonvanishing only when
all of $\varphi_1 (x)$, $\varphi_2 (x)$, and $\varphi_3 (x)$ are in $\mathcal{H}_1$,
so the sign factor $(-1)^{{\rm deg} (\varphi_1)}$ is trivial, but it can be nontrivial
for more general $A_\infty$ algebras.
Using $\omega$, $Q$, and $b_2$, 
the action $S^{(0)}$ can be written
for $\varphi (x)$ in $\mathcal{H}_1$ as follows:
\begin{equation}
S^{(0)} = \frac{1}{2} \, \omega \, ( \, \varphi (x), Q \, \varphi (x) \, )
+\frac{1}{3} \, \omega \, ( \, \varphi (x) \,, b_2 \, ( \, \varphi (x) \otimes \varphi (x) \, ) \, ) \,. 
\end{equation}

Here we should comment on how we treat the commutative nature of the cubic interaction.
While the integral
\begin{equation}
\int d^d x \, \varphi_1 (x) \, \varphi_2 (x) \,\varphi_3 (x)  
\end{equation}
is totally symmetric,
\begin{equation}
\begin{split}
& \int d^d x \, \varphi_1 (x) \, \varphi_2 (x) \,\varphi_3 (x) 
= \int d^d x \, \varphi_1 (x) \, \varphi_3 (x) \,\varphi_2 (x) 
= \int d^d x \, \varphi_2 (x) \, \varphi_1 (x) \,\varphi_3 (x) \\
& = \int d^d x \, \varphi_2 (x) \, \varphi_3 (x) \,\varphi_1 (x) 
= \int d^d x \, \varphi_3 (x) \, \varphi_1 (x) \,\varphi_2 (x)
= \int d^d x \, \varphi_3 (x) \, \varphi_2 (x) \,\varphi_1 (x) \,,
\end{split}
\label{totally-symmetric}
\end{equation}
we only use the cyclic property~\eqref{cubic-cyclic}
when we describe the theory in terms of an $A_\infty$ algebra.
The symmetric property~\eqref{totally-symmetric}
is a consequence of the fact that
the product~\eqref{b_2} is commutative,
\begin{equation}
b_2 \, ( \, \varphi_1 (x) \otimes \varphi_2 (x) \, )
= b_2 \, ( \, \varphi_2 (x) \otimes \varphi_1 (x) \, ) \,,
\end{equation}
but this is not always the case for theories described by $A_\infty$ algebras.
Some of the calculations in this paper simplify
if we use this commutative property of the product,
but we {\it never} use it in this paper
because our primary motivation is to generalize the analysis
to open string field theory where the product is not commutative
and to evaluate correlation functions in the $1/N$ expansion.
As long as we distinguish
$b_2 \, ( \, \varphi_1 (x) \otimes \varphi_2 (x) \, )$
and $b_2 \, ( \, \varphi_2 (x) \otimes \varphi_1 (x) \, )$,
we can unambiguously determine
the topology of non-planar diagrams,
which can be seen explicitly
by generalizing $\varphi (x)^3$
to $\varphi_{ij} (x) \, \varphi_{jk} (x) \, \varphi_{ki} (x)$
for a matrix field $\varphi_{ij} (x)$
and writing Feynman diagrams using the double-line notation,
and this is exactly what we do
when we perform the $1/N$ expansion.

For more general interactions,
we introduce $m_n$ which is a degree-odd linear map from $\mathcal{H}^{\otimes n}$ to $\mathcal{H}$
in order to describe terms of $O(\varphi^{n+1})$ in the action, where we denoted the tensor product
of $n$ copies of $\mathcal{H}$ by $\mathcal{H}^{\otimes n}$:
\begin{equation}
\mathcal{H}^{\otimes n}
= \underbrace{\, \mathcal{H} \otimes \ldots \otimes \mathcal{H} \,}_n \,.
\end{equation}
When we consider quantum corrections to $Q$, we use $m_1$
which is a degree-odd linear map from $\mathcal{H}$ to $\mathcal{H}$.
As in the case of $b_2$,
$m_n ( \, \varphi_1 (x) \otimes \ldots \otimes \varphi_n (x) \, )$ can be nonvanishing
only when all of $\varphi_1 (x)$, $\varphi_2 (x)$, \ldots , and $\varphi_n (x)$ are in $\mathcal{H}_1$,
and in this case $m_n ( \, \varphi_1 (x) \otimes \ldots \otimes \varphi_n (x) \, )$ is in $\mathcal{H}_2$.
We summarize this property as follows:
\begin{equation}
m_n: \underbrace{\, \mathcal{H}_1 \otimes \ldots \otimes \mathcal{H}_1 \,}_n \to \mathcal{H}_2 \,.
\end{equation}

We also consider terms linear in $\varphi$ in the action.
To describe such linear terms we introduce a one-dimensional vector space
given by multiplying a single basis vector ${\bf 1}$ by complex numbers
and denote the vector space by $\mathcal{H}^{\otimes 0}$.
The vector {\bf 1} is degree even and satisfies
\begin{equation}
{\bf 1} \otimes \varphi (x) = \varphi (x) \,, \qquad \varphi (x) \otimes {\bf 1} = \varphi (x)
\end{equation}
for any $\varphi (x)$ in $\mathcal{H}$.
We then introduce $m_0$ which is a degree-odd linear map from $\mathcal{H}^{\otimes 0}$ to $\mathcal{H}$
and $m_0 \, {\bf 1}$ is in $\mathcal{H}_2$.

We require the following cyclic property for $m_n$:
\begin{equation}
\begin{split}
& \omega \, ( \, \varphi_1 (x),
m_n \, ( \, \varphi_2 (x) \otimes \ldots \otimes \varphi_{n+1} (x) \, ) \, ) \\
& = {}-(-1)^{{\rm deg} (\varphi_1)} \,
\omega \, ( \, m_n \, ( \, \varphi_1 (x) \otimes \ldots \otimes \varphi_n (x) \, ) \,,
\varphi_{n+1} (x) \, ) \,.
\end{split}
\label{m_n-cyclicity}
\end{equation}
Again in the current case the symplectic form can be nonvanishing only when
all of $\varphi_1 (x)$, $\varphi_2 (x)$, \ldots , and $\varphi_{n+1} (x)$ are in $\mathcal{H}_1$,
so the sign factor $(-1)^{{\rm deg} (\varphi_1)}$ is trivial, but it can be nontrivial
for more general $A_\infty$ algebras.

We consider an action of the form
\begin{equation}
S = \frac{1}{2} \, \omega \, ( \, \varphi (x), Q \, \varphi (x) \, )
+\sum_{n=0}^\infty \, \frac{1}{n+1} \,
\omega \, ( \, \varphi (x) \,, m_n \, ( \, \varphi (x) \otimes \ldots \otimes \varphi (x) \, ) \, )
\end{equation}
for $\varphi (x)$ in $\mathcal{H}_1$.
The action is written in terms of $Q$ and the set of maps $\{ \, m_n \, \}$,
and it is invariant under the gauge transformation which is also written
in terms of $Q$ and $\{ \, m_n \, \}$
when a set of relations called $A_\infty$ relations are satisfied
among $Q$ and $\{ \, m_n \, \}$.
The relation $Q^2= 0$ we mentioned before
is one of the $A_\infty$ relations when $m_0$ and $m_1$ vanish.
Two more examples of $A_\infty$ relations when $m_0$ and $m_1$ vanish are given by
\begin{align}
& Q \, m_2
+m_2 \, ( \, Q \otimes {\mathbb I} +{\mathbb I} \otimes Q \, ) = 0 \,,
\label{second-A_infinity} \\
& Q \, m_3
+m_2 \, ( \, m_2 \otimes {\mathbb I} +{\mathbb I} \otimes m_2 \, )
+m_3 \, ( \, Q \otimes {\mathbb I} \otimes {\mathbb I}
+{\mathbb I} \otimes Q \otimes {\mathbb I}
+{\mathbb I} \otimes {\mathbb I} \otimes Q \, ) = 0 \,,
\label{third-A_infinity}
\end{align}
where ${\mathbb I}$ is the identity operator on $\mathcal{H}$.
In the current case these relations are trivially satisfied
because $m_n ( \, \varphi_1 (x) \otimes \ldots \otimes \varphi_n (x) \, )$ can be nonvanishing
only when all of $\varphi_1 (x)$, $\varphi_2 (x)$, \ldots , and $\varphi_n (x)$ are in $\mathcal{H}_1$
and $m_n ( \, \varphi_1 (x) \otimes \ldots \otimes \varphi_n (x) \, )$ is in $\mathcal{H}_2$.

\subsection{The coalgebra representation}
\label{subsection-2.2}

To describe the $A_\infty$ relations to all orders,
it is convenient to consider linear operators
acting on the vector space $T \mathcal{H}$ defined by
\begin{equation}
T \mathcal{H}
= \mathcal{H}^{\otimes 0} \, \oplus \mathcal{H}
\oplus \mathcal{H}^{\otimes 2} \oplus \mathcal{H}^{\otimes 3} \oplus  \ldots \,.
\end{equation}
We denote the projection operator onto $\mathcal{H}^{\otimes n}$ by $\pi_n$.
For a map $c_n$ from $\mathcal{H}^{\otimes n}$ to $\mathcal{H}$,
we define an associated operator ${\bm c}_n$ acting on $T \mathcal{H}$ as follows.
The action on the sector $\mathcal{H}^{\otimes m}$ vanishes when $m < n$:
\begin{equation}
{\bm c}_n \, \pi_m = 0 \quad \text{for} \quad m < n \,.
\end{equation}
The action on the sector $\mathcal{H}^{\otimes n}$
is given by $c_n$:
\begin{equation}
{\bm c}_n \, \pi_n = c_n \, \pi_n \,.
\end{equation}
The action on the sector $\mathcal{H}^{\otimes n+1}$ is given by
\begin{equation}
{\bm c}_n \, \pi_{n+1}
= ( \, c_n \otimes {\mathbb I} +{\mathbb I} \otimes c_n \, ) \, \pi_{n+1} \,.
\end{equation}
The action on the sector $\mathcal{H}^{\otimes m}$
for $m > n+1$ is given by
\begin{equation}
{\bm c}_n \, \pi_m = \Bigl( \, c_n \otimes {\mathbb I}^{\otimes (m-n)}
+\sum_{k=1}^{m-n-1} {\mathbb I}^{\otimes k} \otimes c_n \otimes {\mathbb I}^{\otimes (m-n-k)}
+{\mathbb I}^{\otimes (m-n)} \otimes c_n \, \Bigr) \, \pi_m
\quad \text{for} \quad m > n+1 \,,
\label{c_n-pi_m}
\end{equation}
where
\begin{equation}
{\mathbb I}^{\otimes k} = \underbrace{ \, {\mathbb I} \otimes {\mathbb I} \otimes \ldots
\otimes {\mathbb I} \, }_{k} \,.
\end{equation}
An operator acting on $T \mathcal{H}$ of this form is called a {\it coderivation}.\footnote{
Coderivations can be characterized using the coproduct.
See, for example, appendix~A of~\cite{Erler:2015uba} for details.
}
It will be helpful to explain 
the action of ${\bm c}_0$ in more detail.
For example, the action of ${\bm c}_0$ on $\mathcal{H}$ is given by
\begin{equation}
{\bm c}_0 \, \pi_1
= ( \, c_0 \otimes {\mathbb I} +{\mathbb I} \otimes c_0 \, ) \, \pi_1 \,,
\end{equation}
but this should be understood as follows.
Since $\varphi (x)$ in $\mathcal{H}$ can be written as
\begin{equation}
\varphi (x) = {\bf 1} \otimes \varphi (x)
\end{equation}
or as
\begin{equation}
\varphi (x) = \varphi (x) \otimes {\bf 1} \,,
\end{equation}
the action of ${\bm c}_0$ on $\varphi (x)$ should be understood as
\begin{equation}
\begin{split}
{\bm c}_0 \, \varphi (x)
& = ( \, c_0 \otimes {\mathbb I} +{\mathbb I} \otimes c_0 \, ) \, \varphi (x) \\
& = ( \, c_0 \otimes {\mathbb I} \, ) \, ( \, {\bf 1} \otimes \varphi (x) \, )
+( \, {\mathbb I} \otimes c_0 \, ) \, ( \, \varphi (x) \otimes {\bf 1} \, ) \\
& = c_0 \, {\bf 1} \otimes \varphi (x)
+(-1)^{\rm deg (c_0) \, \rm deg (\varphi)}\varphi (x) \otimes c_0 \, {\bf 1} \,.
\end{split}
\end{equation}

We define ${\bm m}$ by
\begin{equation}
{\bm m} = \sum_{n=0}^\infty {\bm m}_n \,,
\end{equation}
where ${\bm m}_n$ is the coderivation associated with $m_n$,
and the $A_\infty$ relations can be written compactly as
\begin{equation}
( \, {\bf Q} +{\bm m} \, )^2 = 0 \,,
\label{A_infinity}
\end{equation}
where ${\bf Q}$ is the coderivation associated with $Q$.
When ${\bm m}_0$ and ${\bm m}_1$ vanish,
the nilpotency of $Q$ in~\eqref{nilpotency} is reproduced 
by the condition $\pi_1 \, ( \, {\bf Q} +{\bm m} \, )^2 \, \pi_1 = 0$
and the relations~\eqref{second-A_infinity} and~\eqref{third-A_infinity}
are reproduced by the conditions $\pi_1 \, ( \, {\bf Q} +{\bm m} \, )^2 \, \pi_2 = 0$
and $\pi_1 \, ( \, {\bf Q} +{\bm m} \, )^2 \, \pi_3 = 0$, respectively.
When a coderivation ${\bm m}$ is given,
we can uniquely determine $m_n$ by the decomposition
\begin{equation}
\pi_1 \, {\bm m} = \sum_{n=0}^\infty m_n \, \pi_n \,.
\label{extraction}
\end{equation}
Therefore, the construction of an action with an $A_\infty$ structure
amounts to the construction of a degree-odd coderivation ${\bm m}$
which satisfies~\eqref{A_infinity}.

\subsection{Projections}
\label{subsection-2.3}

As we wrote in the introduction, homotopy algebras are useful
when we consider projections onto subspaces of $\mathcal{H}$.
We consider projections which commute with $Q$,
and we denote the projection operator by $P$.
It satisfies the following relations:
\begin{equation}
P^2 = P \,, \qquad
Q \, P = P \, Q \,.
\label{P-relations}
\end{equation}
An important ingredient is $h$ which is a degree-odd linear operator on $\mathcal{H}$
and satisfies the following relations:
\begin{equation}
Q \, h +h \, Q = {\mathbb I}-P \,, \qquad
h \, P = 0 \,, \qquad
P \, h = 0 \,, \qquad
h^2 = 0 \,.
\label{h-relations}
\end{equation}
We then promote $P$ and $h$ to the linear operators ${\bf P}$ and ${\bm h}$ on $T \mathcal{H}$, respectively.
The operator ${\bf P}$ is defined by
\begin{equation}
{\bf P} = \pi_0 +\sum_{n=1}^\infty P^{\otimes n} \, \pi_n \,,
\end{equation}
where
\begin{equation}
P^{\otimes n} = \underbrace{ \, P \otimes P \otimes \ldots \otimes P \, }_{n} \,.
\end{equation}
The operator ${\bm h}$ is defined as follows.
Its action on $\mathcal{H}^{\otimes 0}$ vanishes:
\begin{equation}
{\bm h} \, \pi_0 = 0 \,.
\end{equation}
The action on $\mathcal{H}$ is given by $h$:
\begin{equation}
{\bm h} \, \pi_1 = h \, \pi_1 \,.
\end{equation}
The action on $\mathcal{H} \otimes \mathcal{H}$ is given by
\begin{equation}
{\bm h} \, \pi_2
= ( \, h \otimes P +{\mathbb I} \otimes h \, ) \, \pi_2 \,.
\end{equation}
The action on $\mathcal{H}^{\otimes n}$
for $n > 2$ is given by
\begin{equation}
{\bm h} \, \pi_n = \Bigl( \, h \otimes P^{\otimes (n-1)}
+\sum_{m=1}^{n-2} {\mathbb I}^{\otimes m} \otimes h \otimes P^{\otimes (n-m-1)}
+{\mathbb I}^{\otimes (n-1)} \otimes h \, \Bigr) \, \pi_n
\quad \text{for} \quad n > 2 \,.
\label{boldface-h}
\end{equation}
Unlike the case of coderivations, the projection operator $P$ appears
in the definition of ${\bm h}$.
Note also that the appearance of $P$ is asymmetric and
the operator $P$ always appears to the right of $h$.
This property of ${\bm h}$ will play an important role later.
The relations in~\eqref{P-relations} are promoted
to the following relations for ${\bf P}$ and ${\bf Q}$:
\begin{equation}
{\bf P}^2 = {\bf P} \,, \qquad
{\bf Q} \, {\bf P} = {\bf P} \, {\bf Q} \,.
\end{equation}
The relations in~\eqref{h-relations} are promoted
to the following relations involving ${\bm h}$:
\begin{equation}
{\bf Q} \, {\bm h} +{\bm h} \, {\bf Q} = {\bf I}-{\bf P} \,, \qquad
{\bm h} \, {\bf P} = 0 \,, \qquad
{\bf P} \, {\bm h} = 0 \,, \qquad
{\bm h}^2 = 0 \,,
\end{equation}
where ${\bf I}$ is the identity operator on $T \mathcal{H}$.

When the classical action is described by the coderivation given by
\begin{equation}
{\bf Q} +{\bm m}^{(0)} \,,
\end{equation}
the action of the theory projected onto a subspace of $\mathcal{H}$ is
described by
\begin{equation}
{\bf P} \, {\bf Q} \, {\bf P}
+{\bf P} \, {\bm m}^{(0)} \, {\bm f}^{(0)} \, {\bf P}
\label{projected-theory}
\end{equation}
with
\begin{equation}
{\bm f}^{(0)} = \frac{1}{{\bf I} +{\bm h} \, {\bm m}^{(0)}} \,,
\end{equation}
where the inverse of ${\bf I} +{\bm h} \, {\bm m}^{(0)}$ is defined by
\begin{equation}
\frac{1}{{\bf I} +{\bm h} \, {\bm m}^{(0)}}
= {\bf I} +\sum_{n=1}^\infty \, (-1)^n \, ( \, {\bm h} \, {\bm m}^{(0)} \, )^n \,.
\end{equation}
The general construction of~\eqref{projected-theory} from ${\bf Q} +{\bm m}^{(0)}$
is known as the homological perturbation lemma,
and it is described in detail
including the case where $m^{(0)}_0$ is nonvanishing in~\cite{Koyama:2020qfb}.
The action of the theory projected onto a subspace of $\mathcal{H}$
can be constructed from~\eqref{projected-theory}
via the decomposition analogous to~\eqref{extraction}.

When we consider on-shell scattering amplitudes,
we use the projection onto on-shell states.
In that case ${\bf P} \, {\bf Q} \, {\bf P}$ vanishes,
and on-shell scattering amplitudes at the tree level can be calculated from
\begin{equation}
{\bf P} \, {\bm m}^{(0)} \, {\bm f}^{(0)} \, {\bf P} \,.
\end{equation}
When the action including counterterms is described by the coderivation given by
\begin{equation}
{\bf Q} +{\bm m} \,,
\end{equation}
on-shell scattering amplitudes including loop diagrams can be calculated from
\begin{equation}
{\bf P} \, {\bm m} \, {\bm f} \, {\bf P}
\end{equation}
with
\begin{equation}
{\bm f} = \frac{1}{{\bf I} +{\bm h} \, {\bm m} +i \hbar \, {\bm h} \, {\bf U}} \,,
\end{equation}
where the inverse of ${\bf I} +{\bm h} \, {\bm m} +i \hbar \, {\bm h} \, {\bf U}$ is defined by
\begin{equation}
\frac{1}{{\bf I} +{\bm h} \, {\bm m} +i \hbar \, {\bm h} \, {\bf U}}
= {\bf I} +\sum_{n=1}^\infty \,
(-1)^n \, ( \, {\bm h} \, {\bm m} +i \hbar \, {\bm h} \, {\bf U} \, )^n \,.
\end{equation}
The operator ${\bf U}$ consists of maps from $\mathcal{H}^{\otimes n}$ to $\mathcal{H}^{\otimes (n+2)}$.
When the vector space $\mathcal{H}$ is given by $\mathcal{H}_1 \oplus \mathcal{H}_2$,
the operator ${\bf U}$ incorporates a pair of basis vectors of $\mathcal{H}_1$ and $\mathcal{H}_2$.
We denote the basis vector of $\mathcal{H}_1$ by $e^\alpha$,
where $\alpha$ is the label of the basis vectors.
For $\mathcal{H}_2$ we denote the basis vector by $e_\alpha$,
and repeated indices are implicitly summed over.
These basis vectors are normalized as follows:
\begin{equation}
\omega \, ( \, e^\alpha, e_\beta \, ) \, e^\beta = e^\alpha \,, \qquad 
e_\alpha \, \omega \, ( \, e^\alpha, e_\beta \, ) = e_\beta \,. 
\end{equation}
In this paper, we choose $e^\alpha$ and $e_\alpha$ which appear in $T \mathcal{H}$ as
\begin{equation}
\begin{split}
& \ldots \otimes e^\alpha \otimes \ldots \otimes e_\alpha \otimes \ldots
= \int \frac{d^d p}{(2 \pi)^d} \,
\ldots \otimes e^{-ipx} \otimes \ldots \otimes e^{ipx} \otimes \ldots \,.
\end{split}
\end{equation}
The action of ${\bf U}$ on $\mathcal{H}^{\otimes 0}$ is given by
\begin{equation}
{\bf U} \, {\bf 1} = e_\alpha \otimes e^\alpha +e^\alpha \otimes e_\alpha \,,
\end{equation}
and the action of ${\bf U}$ on $\mathcal{H}$ is given by
\begin{equation}
\begin{split}
{\bf U} \, \varphi (x)
& = e_\alpha \otimes e^\alpha \otimes \varphi (x)
+e^\alpha \otimes e_\alpha \otimes \varphi (x)
+(-1)^{\rm deg (\varphi)}
e^\alpha \otimes \varphi (x) \otimes e_\alpha \\
& \quad~ +e_\alpha \otimes \varphi (x) \otimes e^\alpha
+(-1)^{\rm deg (\varphi)}
\varphi (x) \otimes e_\alpha \otimes e^\alpha
+(-1)^{\rm deg (\varphi)}
\varphi (x) \otimes e^\alpha \otimes e_\alpha \,.
\end{split}
\end{equation}
The expressions of ${\bf U} \, {\bf 1}$ and ${\bf U} \, \varphi (x)$
are graded symmetric, but this is not generically the case
when ${\bf U}$ acts on $\mathcal{H}^{\otimes n}$ with $n \ge 2$.
The action of ${\bf U}$ on $\mathcal{H}^{\otimes 2}$ is given by
\begin{equation}
\begin{split}
& {\bf U} \, ( \, \varphi_1 (x) \otimes \varphi_2 (x) \, )
= e_\alpha \otimes e^\alpha \otimes \varphi_1 (x) \otimes \varphi_2 (x)
+e^\alpha \otimes e_\alpha \otimes \varphi_1 (x) \otimes \varphi_2 (x) \\
& \quad~  +(-1)^{\rm deg (\varphi_1)}
e^\alpha \otimes \varphi_1 (x) \otimes e_\alpha \otimes \varphi_2 (x)
+(-1)^{\rm deg (\varphi_1)+\rm deg (\varphi_2)}
e^\alpha \otimes \varphi_1 (x) \otimes \varphi_2 (x) \otimes e_\alpha \\
& \quad~  +e_\alpha \otimes \varphi_1 (x) \otimes e^\alpha \otimes \varphi_2 (x)
+(-1)^{\rm deg (\varphi_1)}
\varphi_1 (x) \otimes e_\alpha \otimes e^\alpha \otimes \varphi_2 (x) \\
& \quad~  +(-1)^{\rm deg (\varphi_1)}
\varphi_1 (x) \otimes e^\alpha \otimes e_\alpha \otimes \varphi_2 (x)
+(-1)^{\rm deg (\varphi_1)+\rm deg (\varphi_2)}
\varphi_1 (x) \otimes e^\alpha \otimes \varphi_2 (x) \otimes e_\alpha \\
& \quad~  +e_\alpha \otimes \varphi_1 (x) \otimes \varphi_2 (x) \otimes e^\alpha 
+(-1)^{\rm deg (\varphi_1)}
\varphi_1 (x) \otimes e_\alpha \otimes \varphi_2 (x) \otimes e^\alpha \\ 
& \quad~ +(-1)^{\rm deg (\varphi_1)+\rm deg (\varphi_2)}
\varphi_1 (x) \otimes \varphi_2 (x) \otimes e_\alpha \otimes e^\alpha 
+(-1)^{\rm deg (\varphi_1)+\rm deg (\varphi_2)}
\varphi_1 (x) \otimes \varphi_2 (x) \otimes e^\alpha \otimes e_\alpha \,.
\end{split}
\end{equation}
In this paper the operator ${\bf U}$ only appears
in the combination ${\bm h} \, {\bf U}$
and we will later present a precise form
of ${\bm h} \, {\bf U}$
when it acts on a space which is relevant to the analysis in this paper.

\subsection{Formula for correlation functions}
\label{subsection-2.4}

In the case of scalar field theories in Euclidean space,
the equation of motion of the free theory is given by
\begin{equation}
Q \, \varphi (x) = ( \, {}-\partial^2 +m^2 \, ) \, \varphi (x) = 0
\end{equation}
for $\varphi (x)$ in $\mathcal{H}_1$.
The solution is unique and is given by
\begin{equation}
\varphi (x) = 0 \,.
\end{equation}
The projection onto the cohomology of $Q$ defines a minimal model
and plays an important role in homotopy algebras.
In the current case,
the projection onto the cohomology of $Q$ corresponds to the projection operator $P$ given by
\begin{equation}
P = 0 \,,
\end{equation}
and the associated operator ${\bf P}$ corresponds to the projection onto $\mathcal{H}^{\otimes 0}$:
\begin{equation}
{\bf P} = \pi_0 \,.
\end{equation} 
The operator ${\bf P} \, {\bm m} \, {\bm f} \, {\bf P}$ vanishes,
and we may consider that the theory is trivial.
However, the operator ${\bm f}$ is nonvanishing
and this operator plays a central role in generating Feynman diagrams.
In the case of the theory in Euclidean space,\footnote{
Unlike the Minkowski case
we do not write $\hbar$ explicitly for the Euclidean case
because theories in Euclidean space can also be regarded as canonical ensembles
of classical statistical mechanics and we can consider them in broader contexts.
If we prefer, we can replace ${\bm h} \, {\bf U}$ with $\hbar \, {\bm h} \, {\bf U}$
or with $\beta^{\, -1} {\bm h} \, {\bf U}$.
}
the appropriate definition of ${\bm f}$ is
\begin{equation}
{\bm f} = \frac{1}{{\bf I} +{\bm h} \, {\bm m} -{\bm h} \, {\bf U}} \,,
\end{equation}
where the inverse of ${\bf I} +{\bm h} \, {\bm m} -{\bm h} \, {\bf U}$ is defined by
\begin{equation}
\frac{1}{{\bf I} +{\bm h} \, {\bm m} -{\bm h} \, {\bf U}}
= {\bf I} +\sum_{n=1}^\infty \,
(-1)^n \, ( \, {\bm h} \, {\bm m} -{\bm h} \, {\bf U} \, )^n \,.
\end{equation}

What does the projection with $P=0$ mean?
If we recall that the projection onto the massless sector
discussed in~\cite{Sen:2016qap, Erbin:2020eyc, Koyama:2020qfb, Arvanitakis:2020rrk, Arvanitakis:2021ecw}
corresponds to integrating out massive fields,
the projection with $P=0$ should correspond
to carrying out the path integral {\it completely}.
This may result in a trivial theory for the classical case,
but it can be nontrivial for the quantum case
and in fact it is exactly what we do when we calculate correlation functions.
We claim that information on correlation functions is encoded
in ${\bm f} \, {\bf 1}$
associated with the case where $P=0$.
More explicitly, correlation functions are given by
\begin{equation}
\langle \, \varphi (x_1) \, \varphi (x_2) \, \ldots \, \varphi (x_n) \, \rangle
= \omega_n \, ( \, \pi_n \, {\bm f} \, {\bf 1} \,,
\delta^d (x-x_1) \otimes \delta^d (x-x_2) \otimes \ldots \otimes \delta^d (x-x_n) \, ) \,,
\label{formula}
\end{equation}
where
\begin{equation}
\omega_n \, ( \, \varphi_1 (x) \otimes \varphi_2 (x) \otimes \ldots \otimes \varphi_n (x) \,,
\varphi'_1 (x) \otimes \varphi'_2 (x) \otimes \ldots \otimes \varphi'_n (x) \, )
= \prod_{i=1}^n \omega \, ( \, \varphi_i (x) \,, \varphi'_i (x) \, ) \,.
\end{equation}
The formula may look complicated, but it states
that $\pi_n \, {\bm f} \, {\bf 1}$ gives the $n$-point function
by simply replacing $x$ with $x_i$ in the $i$-th sector in $\mathcal{H}^{\otimes n}$.
For example, when $\pi_3 \, {\bm f} \, {\bf 1}$ takes the form
\begin{equation}
\pi_3 \, {\bm f} \, {\bf 1}
= \sum_a f_a (x) \otimes g_a (x) \otimes h_a (x) \,,
\end{equation}
the three-point function is given by
\begin{equation}
\begin{split}
& \langle \, \varphi (x_1) \, \varphi (x_2) \, \varphi (x_3) \, \rangle \\
& = \omega_3 \, ( \, \pi_3 \, {\bm f} \, {\bf 1} \,,
\delta^d (x-x_1) \otimes \delta^d (x-x_2) \otimes \delta^d (x-x_3) \, ) \\
& = \sum_a \omega_3 \, ( \, f_a (x) \otimes g_a (x) \otimes h_a (x) \,,
\delta^d (x-x_1) \otimes \delta^d (x-x_2) \otimes \delta^d (x-x_3) \, ) \\
& = \sum_a \omega \, ( \, f_a (x) \,, \delta^d (x-x_1) \, ) \,\,
\omega \, ( \, g_a (x) \,, \delta^d (x-x_2) \, ) \,\,
\omega \, ( \, h_a (x) \,, \delta^d (x-x_3) \, ) \\
& = \sum_a \int d^d x'_1 \, f_a (x'_1) \, \delta^d (x'_1-x_1) \,
\int d^d x'_2 \, g_a (x'_2) \, \delta^d (x'_2-x_2) \,
\int d^d x'_3 \, h_a (x'_3) \, \delta^d (x'_3-x_3) \\
& = \sum_a f_a (x_1) \, g_a (x_2) \, h_a (x_3) \,.
\end{split}
\end{equation}
This can be summarized as the following replacement rule:
\begin{equation}
\begin{split}
\pi_3 \, {\bm f} \, {\bf 1}
& = \sum_a f_a (x) \otimes g_a (x) \otimes h_a (x) \\
& \, \downarrow \\
\langle \, \varphi (x_1) \, \varphi (x_2) \, \varphi (x_3) \, \rangle
& = \sum_a f_a (x_1) \, g_a (x_2) \, h_a (x_3) \,.
\end{split}
\label{replacement}
\end{equation}

We need to construct ${\bm h}$ for the case $P=0$.
The first step is the construction of $h$ satisfying~\eqref{h-relations}.
As $P$ vanishes, the conditions for $h$ are given by
\begin{equation}
Q \, h +h \, Q = {\mathbb I} \,, \qquad
h^2 = 0 \,.
\end{equation}
It is easy to construct $h$ satisfying these equations.
The action of $h$ on $\varphi (x)$ in $\mathcal{H}_2$ is given by
\begin{equation}
h \, \varphi (x) = \int d^d y \int \frac{d^d p}{(2 \pi)^d} \,
\frac{e^{-ip \, (x-y)}}{p^2+m^2} \, \varphi (y) \,,
\end{equation}
and $h \, \varphi (x)$ is in $\mathcal{H}_1$.
On the other hand, the operator $h$ annihilates any element in $\mathcal{H}_1$.
Thus the nonvanishing part of $h$ can be described as follows:
\begin{equation}
h: \mathcal{H}_2 \to \mathcal{H}_1 \,.
\end{equation}
The operator ${\bm h}$ on $T \mathcal{H}$ in the case of $P=0$ is then given by
\begin{equation}
{\bm h} = h \, \pi_1
+\sum_{n=2}^\infty ( \, {\mathbb I}^{\otimes (n-1)} \otimes h \, ) \, \pi_n \,.
\end{equation}

\section{The free theory}
\label{section-3}
\setcounter{equation}{0}

Let us first demonstrate that correlation functions of the free theory
are correctly reproduced.
We denote correlation functions of the free theory by
$\langle \, \varphi (x_1) \, \varphi (x_2) \, \ldots \, \varphi (x_n) \, \rangle^{(0)}$.
In this case the coderivation ${\bm m}$ vanishes
and ${\bm f} \, {\bf 1}$ is given by
\begin{equation}
{\bm f} \, {\bf 1} = \frac{1}{{\bf I} -{\bm h} \, {\bf U}} \, {\bf 1} \,.
\end{equation}

Let us examine the action of the operator ${\bm h} \, {\bf U}$.
It is useful to consider the tensor product of $n$ copies of $\mathcal{H}_1$
and denote it by $\mathcal{H}_1^{\otimes n}$:
\begin{equation}
\mathcal{H}_1^{\otimes n}
= \underbrace{\, \mathcal{H}_1 \otimes \ldots \otimes \mathcal{H}_1 \,}_n \,.
\end{equation}
We also define the vector space $T \mathcal{H}_1$ by
\begin{equation}
T \mathcal{H}_1
= \mathcal{H}^{\otimes 0} \, \oplus \mathcal{H}_1
\oplus \mathcal{H}_1^{\otimes 2} \oplus \mathcal{H}_1^{\otimes 3} \oplus  \ldots \,.
\end{equation}
When ${\bf U}$ acts on $\mathcal{H}_1^{\otimes n}$,
$e_\alpha$ is the only ingredient in $\mathcal{H}_2$
for the resulting element in $\mathcal{H}^{\otimes (n+2)}$.
The following action of ${\bm h}$ can be nonvanishing
only when $h$ in ${\bm h}$ acts on $e_\alpha$.
Therefore, the action of ${\bm h} \, {\bf U}$ on $T \mathcal{H}_1$
is given by
\begin{equation}
\begin{split}
{\bm h} \, {\bf U} & = ( \, e^\alpha \otimes h \, e_\alpha \, ) \, \pi_0
+( \, e^\alpha \otimes {\mathbb I} \otimes h \, e_\alpha
+{\mathbb I} \otimes e^\alpha \otimes h \, e_\alpha \, ) \, \pi_1 \\
& \quad~ +\sum_{n=2}^\infty \,
( \, e^\alpha \otimes {\mathbb I}^{\otimes n} \otimes h \, e_\alpha
+\sum_{m=1}^{n-1} {\mathbb I}^{\otimes m} \otimes e^\alpha \otimes
{\mathbb I}^{\otimes \, (n-m)} \otimes h \, e_\alpha
+{\mathbb I}^{\otimes n} \otimes e^\alpha \otimes h \, e_\alpha \, ) \, \pi_n \,.
\end{split}
\label{hU-action}
\end{equation}
Note that the resulting element is also in $T \mathcal{H}_1$.
Since ${\bf 1}$ is in $T \mathcal{H}_1$,
$( \, {\bm h} \, {\bf U} \, )^n \, {\bf 1}$ is in $T \mathcal{H}_1$
for any~$n$.

When we expand ${\bm f} \, {\bf 1}$ in powers of ${\bm h} \, {\bf U}$,
it is convenient to decompose ${\bm h} \, {\bf U}$ as
\begin{equation}
{\bm h} \, {\bf U} = \sum_{n=0}^\infty \pi_{n+2} \, {\bm h} \, {\bf U} \, \pi_n
= \pi_2 \, {\bm h} \, {\bf U} \, \pi_0
+\pi_3 \, {\bm h} \, {\bf U} \, \pi_1
+\pi_4 \, {\bm h} \, {\bf U} \, \pi_2 +\ldots \,.
\label{hU-decomposition}
\end{equation}
Then it immediately follows that $\pi_n \, {\bm f} \, {\bf 1}$ vanishes when $n$ is odd.
We thus find
\begin{equation}
\langle \, \varphi (x_1) \, \varphi (x_2) \, \ldots \, \varphi (x_n) \, \rangle^{(0)} = 0
\end{equation}
when $n$ is odd.

The two-point function can be calculated from $\pi_2 \, {\bm f} \, {\bf 1}$.
We find
\begin{equation}
\pi_2 \, {\bm f} \, {\bf 1}
= \pi_2 \, {\bm h} \, {\bf U} \, {\bf 1}
= e^\alpha \otimes h \, e_\alpha
= \int \frac{d^d p}{(2 \pi)^d} \, e^{-ipx} \otimes \frac{1}{p^2+m^2} \, e^{ipx} \,.
\end{equation}
Following the replacement rule~\eqref{replacement},
the two-point function is given by
\begin{equation}
\begin{split}
\langle \, \varphi (x_1) \, \varphi (x_2) \, \rangle^{(0)}
& = \omega_2 \, ( \, \pi_2 \, {\bm f} \, {\bf 1} \,,
\delta^d (x-x_1) \otimes \delta^d (x-x_2) \, ) \\
& = \int \frac{d^d p}{(2 \pi)^d} \,
\frac{e^{-ip \, (x_1-x_2)}}{p^2+m^2} \,.
\end{split}
\end{equation}

The four-point function can be calculated from $\pi_4 \, {\bm f} \, {\bf 1}$.
It follows from the decomposition of ${\bm h} \, {\bf U}$ in~\eqref{hU-decomposition} that
\begin{equation}
\pi_4 \, {\bm f} \, {\bf 1}
= \pi_4 \, {\bm h} \, {\bf U} \, {\bm h} \, {\bf U} \, {\bf 1} \,.
\end{equation}
Using the action of ${\bm h} \, {\bf U}$ in~\eqref{hU-action} we find
\begin{equation}
\pi_4 \, {\bm h} \, {\bf U} \, {\bm h} \, {\bf U} \, {\bf 1}
= e^\beta \otimes e^\alpha \otimes h \, e_\alpha \otimes h \, e_\beta
+ e^\alpha \otimes e^\beta \otimes h \, e_\alpha \otimes h \, e_\beta
+ e^\alpha \otimes h \, e_\alpha \otimes e^\beta \otimes h \, e_\beta \,.
\label{hUhU1}
\end{equation}
The explicit form of the first term on the right-hand side is given by
\begin{equation}
\begin{split}
& e^\beta \otimes e^\alpha \otimes h \, e_\alpha \otimes h \, e_\beta \\
& = \int \frac{d^d p_1}{(2 \pi)^d} \int \frac{d^d p_2}{(2 \pi)^d} \,
e^{-ip_2 x} \otimes e^{-ip_1 x}
\otimes \frac{1}{p_1^2+m^2} \, e^{ip_1 x}
\otimes \frac{1}{p_2^2+m^2} \, e^{ip_2 x} \,,
\end{split}
\end{equation}
and the contribution to the four-point function is as follows:
\begin{equation}
\begin{split}
& \omega_4 \, ( \, e^\beta \otimes e^\alpha \otimes h \, e_\alpha \otimes h \, e_\beta \,,
\delta^d (x-x_1) \otimes \delta^d (x-x_2) \otimes \delta^d (x-x_3) \otimes \delta^d (x-x_4) \, ) \\
& = \int \frac{d^d p_1}{(2 \pi)^d} \int \frac{d^d p_2}{(2 \pi)^d} \,
e^{-ip_2 x_1} \, e^{-ip_1 x_2} \,
\frac{1}{p_1^2+m^2} \, e^{ip_1 x_3} \,
\frac{1}{p_2^2+m^2} \, e^{ip_2 x_4} \\
& = \int \frac{d^d p_1}{(2 \pi)^d} \, \frac{e^{-ip_1 \, (x_2-x_3)}}{p_1^2+m^2}
\int \frac{d^d p_2}{(2 \pi)^d} \, \frac{e^{-ip_2 \, (x_1-x_4)}}{p_2^2+m^2} \\
& = \langle \, \varphi (x_2) \, \varphi (x_3) \, \rangle^{(0)} \,
\langle \, \varphi (x_1) \, \varphi (x_4) \, \rangle^{(0)} \,.
\end{split}
\end{equation}
The second and third terms on the right-hand side of~\eqref{hUhU1}
can be calculated similarly,
and the four-point function is given by
\begin{equation}
\begin{split}
& \langle \, \varphi (x_1) \, \varphi (x_2) \, \varphi (x_3) \, \varphi (x_4) \, \rangle^{(0)} \\
& = \omega_4 \, ( \, \pi_4 \, {\bm f} \, {\bf 1} \,,
\delta^d (x-x_1) \otimes \delta^d (x-x_2) \otimes \delta^d (x-x_3) \otimes \delta^d (x-x_4) \, ) \\
& = \langle \, \varphi (x_2) \, \varphi (x_3) \, \rangle^{(0)} \,
\langle \, \varphi (x_1) \, \varphi (x_4) \, \rangle^{(0)}
+\langle \, \varphi (x_1) \, \varphi (x_3) \, \rangle^{(0)} \,
\langle \, \varphi (x_2) \, \varphi (x_4) \, \rangle^{(0)} \\
& \quad~ +\langle \, \varphi (x_1) \, \varphi (x_2) \, \rangle^{(0)} \,
\langle \, \varphi (x_3) \, \varphi (x_4) \, \rangle^{(0)} \,.
\end{split}
\end{equation}
We have thus reproduced Wick's theorem for four-point functions.
It is not difficult to extend the analysis to six-point functions and further,
where Wick's theorem follows from the structure of ${\bm h} \, {\bf U}$ in~\eqref{hU-action}.

\section{$\varphi^3$ theory}
\label{section-4}
\setcounter{equation}{0}

Let us next consider $\varphi^3$ theory
and calculate correlation functions in perturbation theory.\footnote{
Calculations in this section largely overlap
with those for scattering amplitudes in collaboration with Shibuya~\cite{Okawa-Shibuya}.}
The classical action of $\varphi^3$ theory in Euclidean space is given by\footnote{
We consider the theory in six dimensions for $\varphi^3$ theory to be renormalizable,
although we do not explicitly replace $d$ with $6$ in the expressions.
We mostly follow the conventions of the textbook by Srednicki~\cite{Srednicki:2007qs}
converted to the Euclidean case.
}
\begin{equation}
S^{(0)} = \int d^d x \, \biggl[ \, \frac{1}{2} \, \partial_\mu \varphi (x) \, \partial_\mu \varphi (x)
+\frac{1}{2} \, m^2 \, \varphi (x)^2
-\frac{1}{6} \, g \, \varphi (x)^3 \, \biggr] \,,
\end{equation}
and in subsection~\ref{subsection-2.1} we wrote it in the following form:
\begin{equation}
S^{(0)} = \frac{1}{2} \, \omega \, ( \, \varphi (x), Q \, \varphi (x) \, )
+\frac{1}{3} \, \omega \, ( \, \varphi (x) \,, b_2 \, ( \, \varphi (x) \otimes \varphi (x) \, ) \, ) \,. 
\end{equation}
We consider quantum theory, and we need to add counterterms to the classical action.
The action of $\varphi^3$ theory including counterterms is given by
\begin{equation}
S = \int d^d x \, \biggl[ \,
\frac{1}{2} \, Z_\varphi \, \partial_\mu \varphi (x) \, \partial_\mu \varphi (x)
+\frac{1}{2} \, Z_m \, m^2 \, \varphi (x)^2
-\frac{1}{6} \, Z_g \, g \, \varphi (x)^3
-Y \varphi (x) \, \biggr] \,,
\end{equation}
where $Y$, $Z_\varphi$, $Z_m$, and $Z_g$ are constants.
The operators $m_0$, $m_1$, and $m_2$ for this action are defined by
\begin{align}
m_0 \, {\bf 1} & = {}-Y \,, \\
m_1 \, \varphi (x) & = {}-( \, Z_\varphi-1 \, ) \, \partial^2 \varphi (x)
+( \, Z_m-1 \, ) \, m^2 \, \varphi (x) \,, \\
m_2 \, ( \, \varphi_1 (x) \otimes \varphi_2 (x) \, )
& = {}-\frac{g}{2} \, Z_g \, \varphi_1 (x) \, \varphi_2 (x)
\end{align}
for $\varphi (x)$, $\varphi_1 (x)$, and $\varphi_2 (x)$ in $\mathcal{H}_1$.
The coderivations corresponding to $m_0$, $m_1$, and $m_2$
are denoted by ${\bm m}_0$, ${\bm m}_1$, and ${\bm m}_2$,
and we define ${\bm m}$ by
\begin{equation}
{\bm m} = {\bm m}_0 +{\bm m}_1 +{\bm m}_2 \,. 
\end{equation}
The whole action is described by the coderivation ${\bf Q}+{\bm m}$,
and correlation functions can be calculated from ${\bm f} \, {\bf 1}$ given by
\begin{equation}
{\bm f} \, {\bf 1} = \frac{1}{{\bf I} +{\bm h} \, {\bm m} -{\bm h} \, {\bf U}} \, {\bf 1} \,.
\end{equation}

Let us examine the action of the operator ${\bm h} \, {\bm m}$,
which can be divided into ${\bm h} \, {\bm m}_0$, ${\bm h} \, {\bm m}_1$, and ${\bm h} \, {\bm m}_2$.
When ${\bm m}_0$ acts on $\mathcal{H}_1^{\otimes n}$,
$m_0 \, {\bf 1}$ is the only ingredient in $\mathcal{H}_2$
for the resulting element in $\mathcal{H}^{\otimes (n+1)}$.
The following action of ${\bm h}$ can be nonvanishing
only when $h$ in ${\bm h}$ acts on $m_0 \, {\bf 1}$.
Therefore, the action of ${\bm h} \, {\bm m}_0$ on $T \mathcal{H}_1$
is given by
\begin{equation}
{\bm h} \, {\bm m}_0 = h \, m_0 \, \pi_0
+\sum_{n=1}^\infty \,
( \, {\mathbb I}^{\otimes n} \otimes h \, m_0 \, ) \, \pi_n \,.
\label{hm_0-action}
\end{equation}
Similarly, the actions of ${\bm h} \, {\bm m}_1$ and ${\bm h} \, {\bm m}_2$
on $T \mathcal{H}_1$
are given by
\begin{align}
{\bm h} \, {\bm m}_1 & = h \, m_1 \, \pi_1
+\sum_{n=2}^\infty \,
( \, {\mathbb I}^{\otimes \, (n-1)} \otimes h \, m_1 \, ) \, \pi_n \,,
\label{hm_1-action} \\
{\bm h} \, {\bm m}_2 & = h \, m_2 \, \pi_2
+\sum_{n=3}^\infty \,
( \, {\mathbb I}^{\otimes \, (n-2)} \otimes h \, m_2 \, ) \, \pi_n \,.
\label{hm_2-action}
\end{align}
Note that the resulting element is also in $T \mathcal{H}_1$
for each of the actions of ${\bm h} \, {\bm m}_0$,
${\bm h} \, {\bm m}_1$ and ${\bm h} \, {\bm m}_2$ on $T \mathcal{H}_1$.
When we expand ${\bm f} \, {\bf 1}$
in powers of ${\bm h} \, {\bm m}$ and ${\bm h} \, {\bf U}$,
each term in the expansion therefore belongs to $T \mathcal{H}_1$.
For this expansion, it is convenient to decompose
${\bm h} \, {\bm m}_0$, ${\bm h} \, {\bm m}_1$, and ${\bm h} \, {\bm m}_2$ as follows:
\begin{equation}
\begin{split}
{\bm h} \, {\bm m}_0 & = \sum_{n=0}^\infty \pi_{n+1} \, {\bm h} \, {\bm m}_0 \, \pi_n
= \pi_1 \, {\bm h} \, {\bm m}_0 \, \pi_0
+\pi_2 \, {\bm h} \, {\bm m}_0 \, \pi_1
+\pi_3 \, {\bm h} \, {\bm m}_0 \, \pi_2
+ \ldots \,, \\
{\bm h} \, {\bm m}_1 & = \sum_{n=1}^\infty \pi_{n} \, {\bm h} \, {\bm m}_1 \, \pi_n
= \pi_1 \, {\bm h} \, {\bm m}_1 \, \pi_1
+\pi_2 \, {\bm h} \, {\bm m}_1 \, \pi_2
+\pi_3 \, {\bm h} \, {\bm m}_1 \, \pi_3
+ \ldots \,, \\
{\bm h} \, {\bm m}_2 & = \sum_{n=2}^\infty \pi_{n-1} \, {\bm h} \, {\bm m}_2 \, \pi_n
= \pi_1 \, {\bm h} \, {\bm m}_2 \, \pi_2
+\pi_2 \, {\bm h} \, {\bm m}_2 \, \pi_3
+\pi_3 \, {\bm h} \, {\bm m}_2 \, \pi_4
+ \ldots \,.
\label{hm-decomposition}
\end{split}
\end{equation}

Let us calculate correlation functions in perturbation theory
with respect to $g$.
We expand $Y$, $Z_\varphi$, $Z_m$, and $Z_g$ in $g$ as follows:
\begin{align}
Y & = g \, Y^{(1)} +O(g^3) \,, \\
Z_\varphi & = 1 +g^2 Z_\varphi^{(1)} +O(g^4) \,, \\
Z_m & = 1 +g^2 Z_m^{(1)} +O(g^4) \,, \\
Z_g & = 1 +g^2 Z_g^{(1)} +O(g^4) \,.
\end{align}
Correspondingly,
we expand ${\bm m}_0$, ${\bm m}_1$, and ${\bm m}_2$ in $g$ as
\begin{equation}
{\bm m}_0 = \sum_{\ell=0}^\infty {\bm m}_0^{(\ell)} \,, \qquad
{\bm m}_1 = \sum_{\ell=0}^\infty {\bm m}_1^{(\ell)} \,, \qquad
{\bm m}_2 = \sum_{\ell=0}^\infty {\bm m}_2^{(\ell)} \,,
\end{equation}
where ${\bm m}_n^{(\ell)}$ is of $O( g^{n-1+2 \ell \,})$.
We also expand ${\bm m}$ in $g$ as
\begin{equation}
{\bm m} = \sum_{\ell=0}^\infty {\bm m}^{(\ell)} \,,
\end{equation}
where
\begin{equation}
{\bm m}^{(\ell)} = {\bm m}_0^{(\ell)} +{\bm m}_1^{(\ell)} +{\bm m}_2^{(\ell)} \,.
\end{equation}
The coderivation ${\bm m}^{(0)}$ describes the interaction of the classical action
and is given by
\begin{equation}
{\bm m}^{(0)} = {\bm b}_2 \,,
\end{equation}
where ${\bm b}_2$ is the coderivation associated with $b_2$.
The coderivation ${\bm m}^{(1)}$ describes counterterms at one loop,
and $m_0^{(1)}$, $m_1^{(1)}$, and $m_2^{(1)}$ are given by 
\begin{align}
m_0^{(1)} \, {\bf 1} & = {}-g \, Y^{(1)} \,, \\
m_1^{(1)} \, \varphi (x) & = {}-g^2 Z_\varphi^{(1)} \, \partial^2 \varphi (x)
+g^2 Z_m^{(1)} \, m^2 \, \varphi (x) \,, \\
m_2^{(1)} \, ( \, \varphi_1 (x) \otimes \varphi_2 (x) \, )
& = {}-\frac{g^3}{2} \, Z_g^{(1)} \, \varphi_1 (x) \, \varphi_2 (x)
\end{align}
for $\varphi (x)$, $\varphi_1 (x)$, and $\varphi_2 (x)$ in $\mathcal{H}_1$.

\subsection{One-point function}

\begin{figure}[t]
\begin{center}
\includegraphics[scale=0.3]{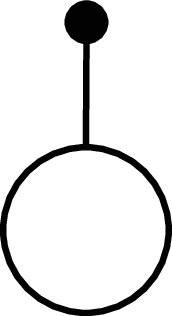}
\end{center}
\caption{One-loop tadpole diagram.
The symmetry factor of this diagram is $2$,
which is correctly reproduced in the calculation of $\langle \, \varphi (x_1) \, \rangle$.
}
\label{figure-1-point}
\end{figure}

The one-point function can be calculated from $\pi_1 \, {\bm f} \, {\bf 1}$.
Let us expand $\pi_1 \, {\bm f} \, {\bf 1}$ in $g$.
Since ${\bm h} \, {\bm m}$ is of $O(g)$, we find
\begin{equation}
\begin{split}
\pi_1 \, {\bm f} \, {\bf 1}
& = \pi_1 \, \frac{1}{{\bf I} +{\bm h} \, {\bm m} -{\bm h} \, {\bf U}} \, {\bf 1} \\
& = \pi_1 \, \frac{1}{{\bf I} -{\bm h} \, {\bf U}} \, {\bf 1}
-\pi_1 \, \frac{1}{{\bf I} -{\bm h} \, {\bf U}} \,  
{\bm h} \, {\bm m} \,
\frac{1}{{\bf I} -{\bm h} \, {\bf U}} \, {\bf 1}  +O(g^2) \,.
\end{split}
\end{equation}
It follows from the decomposition of ${\bm h} \, {\bf U}$ in~\eqref{hU-decomposition} that
\begin{equation}
\pi_1 \, \frac{1}{{\bf I} -{\bm h} \, {\bf U}} = \pi_1 \,.
\end{equation}
Since $\pi_1 \, {\bf 1}$ vanishes, we obtain
\begin{equation}
\pi_1 \, {\bm f} \, {\bf 1}
= {}-\pi_1 \, {\bm h} \, {\bm m} \,
\frac{1}{{\bf I} -{\bm h} \, {\bf U}} \, {\bf 1}  +O(g^2) \,.
\end{equation}
We further use the decomposition of ${\bm h} \, {\bf U}$ in~\eqref{hU-decomposition}
and the decompositions of ${\bm h} \, {\bm m_0}$, ${\bm h} \, {\bm m_1}$, and ${\bm h} \, {\bm m_2}$
in~\eqref{hm-decomposition} to find
\begin{equation}
{}-\pi_1 \, {\bm h} \, {\bm m} \,
\frac{1}{{\bf I} -{\bm h} \, {\bf U}} \, {\bf 1}
= {}-\pi_1 \, {\bm h} \, {\bm m}_2 \, {\bm h} \, {\bf U} \, {\bf 1}
-\pi_1 \, {\bm h} \, {\bm m}_0 \, {\bf 1} \,.
\end{equation}
We then expand ${\bm m}_2$ and ${\bm m}_0$ in $g$ to obtain
\begin{equation}
\begin{split}
\pi_1 \, {\bm f} \, {\bf 1}
& = {}-\pi_1 \, {\bm h} \, {\bm b}_2 \, {\bm h} \, {\bf U} \, {\bf 1}
{}-\pi_1 \, {\bm h} \, {\bm m}_0^{(1)} \, {\bf 1} +O(g^2) \\
& = {}-h \, b_2 \, ( \, e^\alpha \otimes h \, e_\alpha \, )
-h \, m_0^{(1)} \, {\bf 1} +O(g^2) \,.
\end{split}
\end{equation}
The explicit form of the terms of~$O(g)$ is given by
\begin{equation}
{}-h \, b_2 \, ( \, e^\alpha \otimes h \, e_\alpha \, )
-h \, m_0^{(1)} \, {\bf 1}
= \frac{g}{m^2} \, \biggl[ \,
\frac{1}{2} \int \frac{d^d p}{(2 \pi)^d} \, \frac{1}{p^2+m^2}
+Y^{(1)} \, \biggr] \,,
\end{equation}
and the one-point function is given by
\begin{equation}
\begin{split}
\langle \, \varphi (x_1) \, \rangle
& = \omega_1 \, ( \, \pi_1 \, {\bm f} \, {\bf 1} \,,
\delta^d (x-x_1)  \, ) \\
& = \frac{g}{m^2} \, \biggl[ \,
\frac{1}{2} \int \frac{d^d p}{(2 \pi)^d} \, \frac{1}{p^2+m^2}
+Y^{(1)} \, \biggr] +O(g^2) \,.
\end{split}
\end{equation}
We have reproduced the contribution from the one-loop tadpole diagram.
See figure~\ref{figure-1-point}.
Note that the correct symmetry factor appeared.\footnote{
See~\cite{Saemann:2020oyz} for the analysis of symmetry factors
in the context of scattering amplitudes.
}
The integral over $p$ is divergent in six dimensions,
so we need to regularize the integral by introducing a cutoff $\Lambda$.\footnote{
While we can use standard regularizations used in quantum field theory,
it would be interesting to consider 
the non-local deformation discussed in~\cite{Chiaffrino:2021uyd},
which has a structure analogous to string field theory.
}
We then choose the constant $Y^{(1)}$ to depend on $\Lambda$
so that the one-point function at $O(g)$ is finite 
in the limit $\Lambda \to \infty$.
While we can make the one-point function vanish at $O(g)$
by choosing $Y^{(1)}$ to cancel the contribution from the one-loop tadpole diagram,
we leave it finite and keep track of the appearance of one-loop tadpoles.

It is convenient to define $n_0^{(1)}$ by
\begin{equation}
n_0^{(1)}  \, {\bf 1} = b_2 \, ( \, e^\alpha \otimes h \, e_\alpha \, ) \,.
\end{equation}
We can then write $\pi_1 \, {\bm f} \, {\bf 1}$ as follows:
\begin{equation}
\pi_1 \, {\bm f} \, {\bf 1}
= {}-h \, ( \, n_0^{(1)} +m_0^{(1)} \, ) \, {\bf 1} +O(g^2) \,.
\end{equation}
Let us denote the sum $n_0^{(1)} +m_0^{(1)}$ by $\Gamma_0^{(1)}$.
It is given by
\begin{equation}
\Gamma_0^{(1)} \, {\bf 1}
= ( \, n_0^{(1)} +m_0^{(1)} \, ) \, {\bf 1}
= b_2 \, ( \, e^\alpha \otimes h \, e_\alpha \, ) +m_0^{(1)} \, {\bf 1}
= {}-\frac{g}{2} \int \frac{d^d p}{(2 \pi)^d} \, \frac{1}{p^2+m^2}
-g \, Y^{(1)} \,.
\end{equation}
The operator $\Gamma_0^{(1)}$ describes the linear term at one loop
in the one-particle irreducible (1PI) effective action~\cite{Okawa-Shibuya}.
We write the one-point function as
\begin{equation}
\langle \, \varphi (x_1) \, \rangle
= \langle \, \varphi (x_1) \, \rangle^{(1)} +O(g^2) \,,
\end{equation}
where
\begin{equation}
\begin{split}
\langle \, \varphi (x_1) \, \rangle^{(1)}
& = {}-\omega_1 \, ( \, h \, \Gamma_0^{(1)} \, {\bf 1} \,,
\delta^d (x-x_1)  \, ) \\
& = \frac{g}{m^2} \, \biggl[ \,
\frac{1}{2} \int \frac{d^d p}{(2 \pi)^d} \, \frac{1}{p^2+m^2}
+Y^{(1)} \, \biggr] \,.
\end{split}
\end{equation}

\subsection{Two-point function}

The two-point function can be calculated from $\pi_2 \, {\bm f} \, {\bf 1}$.
Let us expand $\pi_2 \, {\bm f} \, {\bf 1}$ in $g$ as follows:
\begin{equation}
\begin{split}
\pi_2 \, {\bm f} \, {\bf 1}
& = \pi_2 \, \frac{1}{{\bf I} +{\bm h} \, {\bm m} -{\bm h} \, {\bf U}} \, {\bf 1} \\
& = \pi_2 \, \frac{1}{{\bf I} -{\bm h} \, {\bf U}} \, {\bf 1}
-\pi_2 \, \frac{1}{{\bf I} -{\bm h} \, {\bf U}} \,  
{\bm h} \, {\bm m} \,
\frac{1}{{\bf I} -{\bm h} \, {\bf U}} \, {\bf 1} \\
& \quad~ +\pi_2 \, \frac{1}{{\bf I} -{\bm h} \, {\bf U}} \,  
{\bm h} \, {\bm m} \,
\frac{1}{{\bf I} -{\bm h} \, {\bf U}} \,
{\bm h} \, {\bm m} \,
\frac{1}{{\bf I} -{\bm h} \, {\bf U}} \, {\bf 1}  +O(g^3) \,.
\end{split}
\label{pi_2-f1}
\end{equation}
It follows from the decomposition of ${\bm h} \, {\bf U}$ in~\eqref{hU-decomposition} that
\begin{equation}
\pi_2 \, \frac{1}{{\bf I} -{\bm h} \, {\bf U}} = \pi_2 +\pi_2 \, {\bm h} \, {\bf U} \, \pi_0 \,.
\label{pi_2-1/(I-hU)}
\end{equation}
When we substitute~\eqref{pi_2-1/(I-hU)} into~\eqref{pi_2-f1},
$\pi_2$ and $\pi_2 \, {\bm h} \, {\bf U} \, \pi_0$
on the right-hand side of~\eqref{pi_2-1/(I-hU)}
act on ${\bf 1}$ or ${\bm h} \, {\bm m}$.
Since $\pi_2 \, {\bf 1}$ and $\pi_0 \, {\bm h} \, {\bm m}$ vanish, we obtain
\begin{equation}
\begin{split}
\pi_2 \, {\bm f} \, {\bf 1}
= \pi_2 \, {\bm h} \, {\bf U} \, {\bf 1}
-\pi_2 \, {\bm h} \, {\bm m} \,
\frac{1}{{\bf I} -{\bm h} \, {\bf U}} \, {\bf 1}
+\pi_2 \, {\bm h} \, {\bm m} \,
\frac{1}{{\bf I} -{\bm h} \, {\bf U}} \,
{\bm h} \, {\bm m} \,
\frac{1}{{\bf I} -{\bm h} \, {\bf U}} \, {\bf 1}  +O(g^3) \,.
\end{split}
\end{equation}
We use the decomposition of ${\bm h} \, {\bf U}$ in~\eqref{hU-decomposition}
and the decompositions of ${\bm h} \, {\bm m_0}$, ${\bm h} \, {\bm m_1}$, and ${\bm h} \, {\bm m_2}$
in~\eqref{hm-decomposition} to find
\begin{equation}
\begin{split}
{}-\pi_2 \, {\bm h} \, {\bm m} \,
\frac{1}{{\bf I} -{\bm h} \, {\bf U}} \, {\bf 1}
& = {}-\pi_2 \, {\bm h} \, {\bm m}_1 \, {\bm h} \, {\bf U} \, {\bf 1} \,, \\
\pi_2 \, {\bm h} \, {\bm m} \,
\frac{1}{{\bf I} -{\bm h} \, {\bf U}} \,
{\bm h} \, {\bm m} \,
\frac{1}{{\bf I} -{\bm h} \, {\bf U}} \, {\bf 1}
& = \pi_2 \, {\bm h} \, {\bm m}_0 \, {\bm h} \, {\bm m}_0 \, {\bf 1}
+\pi_2 \, {\bm h} \, {\bm m}_0 \, {\bm h} \, {\bm m}_2 \, {\bm h} \, {\bf U} \, {\bf 1} \\
& \quad~ +\pi_2 \, {\bm h} \, {\bm m}_1 \, {\bm h} \, {\bm m}_1 \, {\bm h} \, {\bf U} \, {\bf 1} \\
& \quad~ +\pi_2 \, {\bm h} \, {\bm m}_2 \, {\bm h} \, {\bm m}_0 \, {\bm h} \, {\bf U} \, {\bf 1}
+\pi_2 \, {\bm h} \, {\bm m}_2 \, {\bm h} \, {\bm m}_2 \, {\bm h} \,
{\bf U} \, {\bm h} \, {\bf U} \, {\bf 1} \\
& \quad~ +\pi_2 \, {\bm h} \, {\bm m}_2 \, {\bm h} \, {\bf U} \, 
{\bm h} \, {\bm m}_0 \, {\bf 1}
+\pi_2 \, {\bm h} \, {\bm m}_2 \, {\bm h} \, {\bf U} \, 
{\bm h} \, {\bm m}_2 \, {\bm h} \, {\bf U} \,{\bf 1} \,.
\end{split}
\end{equation}
We then expand ${\bm m}_2$, ${\bm m}_1$, and ${\bm m}_0$ in $g$ to obtain
\begin{equation}
\begin{split}
\pi_2 \, {\bm f} \, {\bf 1}
& = \pi_2 \, {\bm h} \, {\bf U} \, {\bf 1}
-\pi_2 \, {\bm h} \, {\bm m}_1^{(1)} \, {\bm h} \, {\bf U} \, {\bf 1} \\
& \quad~ +\pi_2 \, {\bm h} \, {\bm m}_0^{(1)} \, {\bm h} \, {\bm m}_0^{(1)} \, {\bf 1}
+\pi_2 \, {\bm h} \, {\bm m}_0^{(1)} \, {\bm h} \, {\bm b}_2 \, {\bm h} \, {\bf U} \, {\bf 1} \\
& \quad~ +\pi_2 \, {\bm h} \, {\bm b}_2 \, {\bm h} \, {\bm m}_0^{(1)} \, {\bm h} \, {\bf U} \, {\bf 1}
+\pi_2 \, {\bm h} \, {\bm b}_2 \, {\bm h} \, {\bm b}_2 \, {\bm h} \,
{\bf U} \, {\bm h} \, {\bf U} \, {\bf 1} \\
& \quad~ +\pi_2 \, {\bm h} \, {\bm b}_2 \, {\bm h} \, {\bf U} \, 
{\bm h} \, {\bm m}_0^{(1)} \, {\bf 1}
+\pi_2 \, {\bm h} \, {\bm b}_2 \, {\bm h} \, {\bf U} \, 
{\bm h} \, {\bm b}_2 \, {\bm h} \, {\bf U} \,{\bf 1} +O(g^3) \,.
\end{split}
\end{equation}
In addition to the term $\pi_2 \, {\bm h} \, {\bf U} \, {\bf 1}$ of the free theory,
there are seven terms of $O(g^2)$.
Let us first calculate two of them which do not involve counterterms.
One of them is $\pi_2 \, {\bm h} \, {\bm b}_2 \, {\bm h} \, {\bm b}_2 \,
{\bm h} \, {\bf U} \, {\bm h} \, {\bf U} \, {\bf 1}$:
\begin{equation}
\begin{split}
\pi_2 \, {\bm h} \, {\bm b}_2 \, {\bm h} \, {\bm b}_2 \,
{\bm h} \, {\bf U} \, {\bm h} \, {\bf U} \, {\bf 1}
& = e^\beta \otimes h \, b_2 \, ( \, e^\alpha \otimes
h \, b_2 \, ( \, h \, e_\alpha \otimes h \, e_\beta \, ) \, ) \\
& \quad~ + e^\alpha \otimes h \, b_2 \, ( \, e^\beta \otimes
h \, b_2 \, ( \, h \, e_\alpha \otimes h \, e_\beta \, ) \, ) \\
& \quad~ + e^\alpha \otimes h \, b_2 \, ( \, h \, e_\alpha \otimes
h \, b_2 \, ( \, e^\beta \otimes h \, e_\beta \, ) \, ) \\
& = e^\beta \otimes h \, b_2 \, ( \, e^\alpha \otimes
h \, b_2 \, ( \, h \, e_\alpha \otimes h \, e_\beta \, ) \, ) \\
& \quad~ + e^\alpha \otimes h \, b_2 \, ( \, e^\beta \otimes
h \, b_2 \, ( \, h \, e_\alpha \otimes h \, e_\beta \, ) \, ) \\
& \quad~ + e^\alpha \otimes h \, b_2 \, ( \, h \, e_\alpha \otimes h \, n_0^{(1)} \, {\bf 1} \, ) \,.
\end{split}
\end{equation}
The other one is $\pi_2 \, {\bm h} \, {\bm b}_2 \, {\bm h} \, {\bf U} \, 
{\bm h} \, {\bm b}_2 \, {\bm h} \, {\bf U} \,{\bf 1}$:
\begin{equation}
\begin{split}
\pi_2 \, {\bm h} \, {\bm b}_2 \, {\bm h} \, {\bf U} \, 
{\bm h} \, {\bm b}_2 \, {\bm h} \, {\bf U} \,{\bf 1}
& = e^\beta \otimes h \, b_2 \, ( \,
h \, b_2 \, ( \, e^\alpha \otimes h \, e_\alpha \, ) \otimes h \, e_\beta \, ) \\
& \quad~ +h \, b_2 \, ( \, e^\alpha \otimes h \, e_\alpha \, )
\otimes h \, b_2 \, ( \, e^\beta \otimes h \, e_\beta \, ) \\
& = e^\beta \otimes h \, b_2 \, ( \, h \, n_0^{(1)} \, {\bf 1} \otimes h \, e_\beta \, )
+h \, n_0^{(1)} \, {\bf 1} \otimes h \, n_0^{(1)} \, {\bf 1} \,.
\end{split}
\end{equation}
Let us define $n_1^{(1)}$ by
\begin{equation}
n_1^{(1)}
= {}-b_2 \, ( \, e^\alpha \otimes
h \, b_2 \, ( \, {\mathbb I} \otimes h \, e_\alpha +h \, e_\alpha \otimes {\mathbb I} \, ) \, ) \,.
\end{equation}
Then the sum of the two terms is written as
\begin{equation}
\begin{split}
& \pi_2 \, {\bm h} \, {\bm b}_2 \, {\bm h} \, {\bm b}_2 \,
{\bm h} \, {\bf U} \, {\bm h} \, {\bf U} \, {\bf 1}
+\pi_2 \, {\bm h} \, {\bm b}_2 \, {\bm h} \, {\bf U} \, 
{\bm h} \, {\bm b}_2 \, {\bm h} \, {\bf U} \,{\bf 1} \\
& = {}-e^\alpha \otimes h \, n_1^{(1)} \, h \, e_\alpha \\
& \quad~ +e^\alpha \otimes h \, b_2 \, ( \, h \, e_\alpha \otimes h \, n_0^{(1)} \, {\bf 1} \, )
+e^\alpha \otimes h \, b_2 \, ( \, h \, n_0^{(1)} \, {\bf 1} \otimes h \, e_\alpha \, ) \\
& \quad~ +h \, n_0^{(1)} \, {\bf 1} \otimes h \, n_0^{(1)} \, {\bf 1} \,.
\end{split}
\end{equation}
The remaining five terms involving counterterms are given by
\begin{align}
{}-\pi_2 \, {\bm h} \, {\bm m}_1^{(1)} \, {\bm h} \, {\bf U} \, {\bf 1}
& = {}-e^\alpha \otimes h \, m_1^{(1)} h \, e_\alpha \,, \\
\pi_2 \, {\bm h} \, {\bm m}_0^{(1)} \, {\bm h} \, {\bm m}_0^{(1)} \, {\bf 1}
& = h \, m_0^{(1)} \, {\bf 1} \otimes h \, m_0^{(1)} \, {\bf 1} \,, \\
\pi_2 \, {\bm h} \, {\bm m}_0^{(1)} \, {\bm h} \, {\bm b}_2 \, {\bm h} \, {\bf U} \, {\bf 1}
& = h \, b_2 \, ( \, e^\alpha \otimes h \, e_\alpha \, ) \otimes h \, m_0^{(1)} \, {\bf 1}
= h \, n_0^{(1)} \, {\bf 1} \otimes h \, m_0^{(1)} \, {\bf 1} \,, \\
\pi_2 \, {\bm h} \, {\bm b}_2 \, {\bm h} \, {\bm m}_0^{(1)} \, {\bm h} \, {\bf U} \, {\bf 1}
& = e^\alpha \otimes h \, b_2 \, ( \, h \, e_\alpha \otimes h \, m_0^{(1)} \, {\bf 1} \, ) \,, \\
\pi_2 \, {\bm h} \, {\bm b}_2 \, {\bm h} \, {\bf U} \, 
{\bm h} \, {\bm m}_0^{(1)} \, {\bf 1}
& = e^\alpha \otimes h \, b_2 \, ( \, h \, m_0^{(1)} \, {\bf 1} \otimes h \, e_\alpha \, ) 
+h \, m_0^{(1)} \, {\bf 1} \otimes h \, b_2 \, ( \,  e^\alpha \otimes h \, e_\alpha \, )  \nonumber \\
& = e^\alpha \otimes h \, b_2 \, ( \, h \, m_0^{(1)} \, {\bf 1} \otimes h \, e_\alpha \, ) 
+h \, m_0^{(1)} \, {\bf 1} \otimes h \, n_0^{(1)} \, {\bf 1} \,.
\end{align}
We define $\Gamma_1^{(1)}$ by
\begin{equation}
\Gamma_1^{(1)} = n_1^{(1)} +m_1^{(1)} \,.
\end{equation}
Then $\pi_2 \, {\bm f} \, {\bf 1}$ can be written in terms of $\Gamma_1^{(1)}$ and $\Gamma_0^{(1)}$
as follows:
\begin{figure}[hbt]
\begin{center}
\includegraphics[scale=0.3]{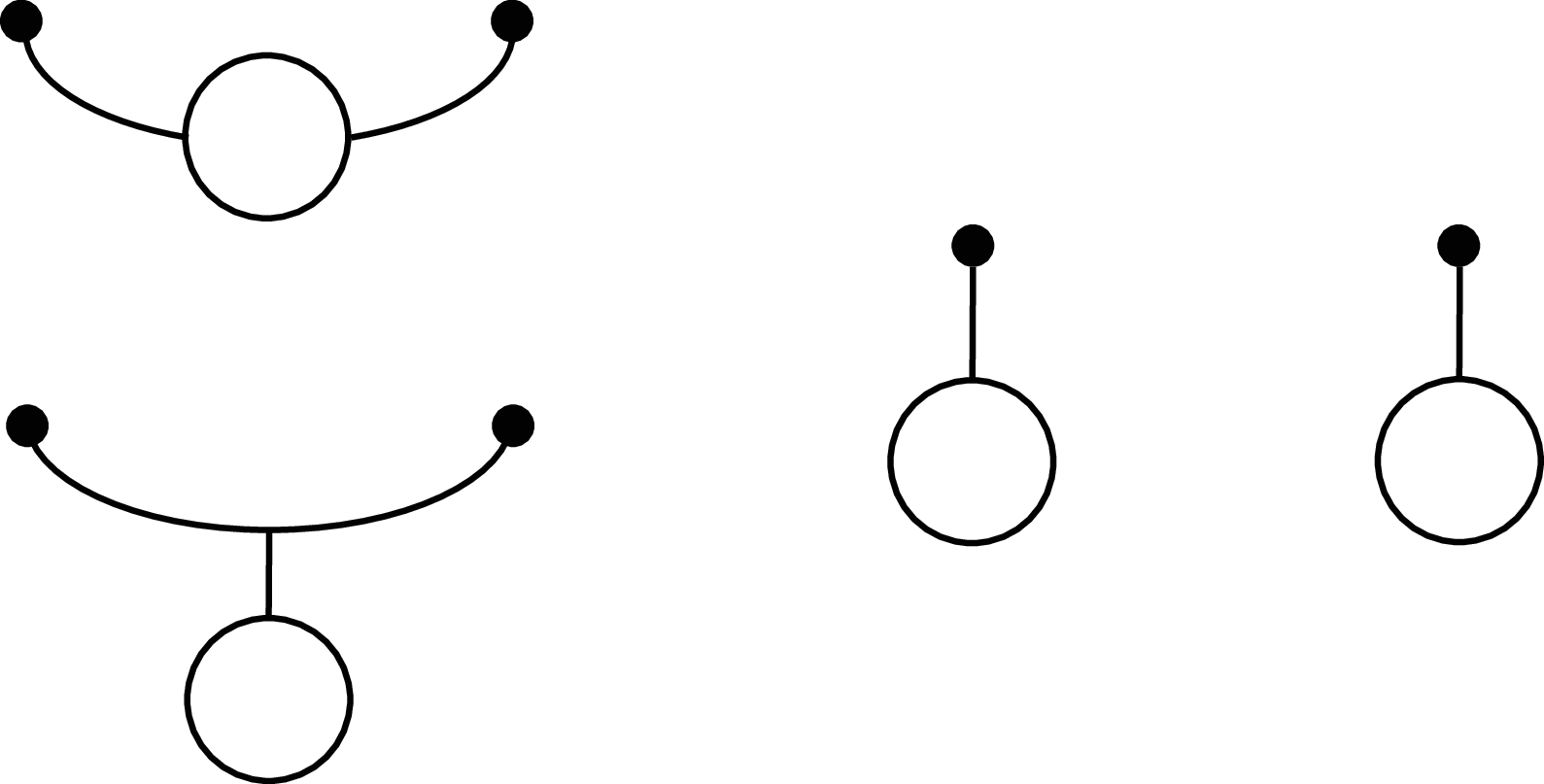}
\end{center}
\caption{Feynman diagrams contributing to the two-point function
$\langle \, \varphi (x_1) \, \varphi (x_2) \, \rangle$ at $O(g^2)$
without involving counterterms.
The top left diagram is associated with $\Gamma_1^{(1)}$.
The bottom left diagram contains the one-loop tadpole.
The disconnected diagrams on the right side
correspond to
$\langle \, \varphi (x_1)  \, \rangle^{(1)} \, \langle \, \varphi (x_2) \, \rangle^{(1)}$.}
\label{figure-2-point}
\end{figure}
\begin{equation}
\begin{split}
\pi_2 \, {\bm f} \, {\bf 1}
& = e^\alpha \otimes h \, e_\alpha
-e^\alpha \otimes h \, \Gamma_1^{(1)} \, h \, e_\alpha \\
& \quad~
+e^\alpha \otimes h \, b_2 \, ( \, h \, e_\alpha \otimes h \, \Gamma_0^{(1)} \, {\bf 1} \, )
+e^\alpha \otimes h \, b_2 \, ( \, h \, \Gamma_0^{(1)} \, {\bf 1} \otimes h \, e_\alpha \, ) \\
& \quad~ +h \, \Gamma_0^{(1)} \, {\bf 1}
\otimes h \, \Gamma_0^{(1)} \, {\bf 1}
+O(g^3) \,.
\end{split}
\end{equation}
The action of $\Gamma_1^{(1)}$ on $e^{ikx}$ in $\mathcal{H}_1$ is given by
\begin{equation}
\Gamma_1^{(1)} \, e^{ikx}
= \biggl[ \, {}-\frac{g^2}{2} \int \frac{d^d p}{(2 \pi)^d} \,
\frac{1}{(p+k)^2+m^2} \, \frac{1}{p^2+m^2}
+g^2 \, Z_\varphi^{(1)} \, k^2
+g^2 \, Z_m^{(1)} \, m^2 \, \biggr] \, e^{ikx} \,.
\end{equation}
The integral over $p$ is divergent in six dimensions,
but we choose the constants $Z_\varphi^{(1)}$ and $Z_m^{(1)}$ to depend on $\Lambda$
so that $\Gamma_1^{(1)}$ is finite in the limit $\Lambda \to \infty$. 
The two-point function is given by
\begin{equation}
\begin{split}
\langle \, \varphi (x_1) \, \varphi (x_2) \, \rangle
& = \omega_2 \, ( \, \pi_2 \, {\bm f} \, {\bf 1} \,,
\delta^d (x-x_1) \otimes \delta^d (x-x_2) \, ) \\
& = \langle \, \varphi (x_1) \, \varphi (x_2) \, \rangle^{(0)}
+\langle \, \varphi (x_1) \, \varphi (x_2) \, \rangle_C^{(1)}
+\langle \, \varphi (x_1)  \, \rangle^{(1)} \, \langle \, \varphi (x_2) \, \rangle^{(1)}
+O(g^3) \,,
\end{split}
\end{equation}
where
\begin{align}
& \langle \, \varphi (x_1) \, \varphi (x_2) \, \rangle_C^{(1)} \nonumber \\
& = {}-\omega_2 \, ( \, 
e^\alpha \otimes h \, \Gamma_1^{(1)} \, h \, e_\alpha \,,
\delta^d (x-x_1) \otimes \delta^d (x-x_2) \, ) \nonumber \\
& \quad~ +\omega_2 \, ( \, 
e^\alpha \otimes h \, b_2 \, ( \, h \, e_\alpha \otimes h \, \Gamma_0^{(1)} \, {\bf 1} \, ) \,,
\delta^d (x-x_1) \otimes \delta^d (x-x_2) \, ) \nonumber \\
& \quad~ +\omega_2 \, ( \, 
e^\alpha \otimes h \, b_2 \, ( \, h \, \Gamma_0^{(1)} \, {\bf 1} \otimes h \, e_\alpha \, ) \,,
\delta^d (x-x_1) \otimes \delta^d (x-x_2) \, ) \nonumber \\
& = {}-g^2 \int \frac{d^d p}{(2 \pi)^d} \, \frac{e^{-ip \, (x_1-x_2)}}{( \, p^2+m^2 \, )^2} \,
\biggl[ \, {}-\frac{1}{2} \int \frac{d^d \ell}{(2 \pi)^d} \,
\frac{1}{(\ell+p)^2+m^2} \, \frac{1}{\ell^2+m^2}
+Z_\varphi^{(1)} \, p^2
+Z_m^{(1)} \, m^2 \, \biggr] \nonumber \\
& \quad~ +g^2 \int \frac{d^d p}{(2 \pi)^d} \, \frac{e^{-ip \, (x_1-x_2)}}{m^2 \, ( \, p^2+m^2 \, )^2} \,
\biggl[ \, \frac{1}{2} \int \frac{d^d \ell}{(2 \pi)^d} \, \frac{1}{\ell^2+m^2}
+Y^{(1)}\, \biggr] \,.
\end{align}
The two-point function $\langle \, \varphi (x_1) \, \varphi (x_2) \, \rangle$ contains
$\langle \, \varphi (x_1)  \, \rangle^{(1)} \, \langle \, \varphi (x_2) \, \rangle^{(1)}$
at $O(g^2)$,
which corresponds to contributions from disconnected Feynman diagrams.
The connected part at one loop
$\langle \, \varphi (x_1) \, \varphi (x_2) \, \rangle_C^{(1)}$
involves $\Gamma_0^{(1)}$ and $\Gamma_1^{(1)}$.
The terms involving $\Gamma_0^{(1)}$ correspond to contributions from
a diagram containing the one-loop tadpole
and a diagram containing the associated counterterm.
The operator $\Gamma_1^{(1)}$ describes the one-loop correction
to the kinetic term in the 1PI effective action~\cite{Okawa-Shibuya}.
See figure~\ref{figure-2-point}.

\subsection{Three-point function}

The three-point function can be calculated from $\pi_3 \, {\bm f} \, {\bf 1}$.
Let us expand $\pi_3 \, {\bm f} \, {\bf 1}$ in $g$ as follows:
\begin{equation}
\begin{split}
\pi_3 \, {\bm f} \, {\bf 1}
& = \pi_3 \, \frac{1}{{\bf I} +{\bm h} \, {\bm m} -{\bm h} \, {\bf U}} \, {\bf 1} \\
& = \pi_3 \, \frac{1}{{\bf I} -{\bm h} \, {\bf U}} \, {\bf 1}
-\pi_3 \, \frac{1}{{\bf I} -{\bm h} \, {\bf U}} \,  
{\bm h} \, {\bm m} \,
\frac{1}{{\bf I} -{\bm h} \, {\bf U}} \, {\bf 1} +O(g^2) \,.
\end{split}
\end{equation}
It follows from the decomposition of ${\bm h} \, {\bf U}$ in~\eqref{hU-decomposition} that
\begin{equation}
\pi_3 \, \frac{1}{{\bf I} -{\bm h} \, {\bf U}} = \pi_3 +\pi_3 \, {\bm h} \, {\bf U} \, \pi_1 \,.
\end{equation}
Since $\pi_3 \, {\bf 1}$ and $\pi_1 \, {\bf 1}$ vanish, we obtain
\begin{equation}
\pi_3 \, {\bm f} \, {\bf 1}
= {}-\pi_3 \, {\bm h} \, {\bm m} \,
\frac{1}{{\bf I} -{\bm h} \, {\bf U}} \, {\bf 1}
-\pi_3 \, {\bm h} \, {\bf U} \, {\bm h} \, {\bm m} \,
\frac{1}{{\bf I} -{\bm h} \, {\bf U}} \, {\bf 1} +O(g^2) \,.
\end{equation}
We use the decomposition of ${\bm h} \, {\bf U}$ in~\eqref{hU-decomposition}
and the decompositions of ${\bm h} \, {\bm m_0}$, ${\bm h} \, {\bm m_1}$, and ${\bm h} \, {\bm m_2}$
in~\eqref{hm-decomposition} to find
\begin{equation}
\begin{split}
{}-\pi_3 \, {\bm h} \, {\bm m} \,
\frac{1}{{\bf I} -{\bm h} \, {\bf U}} \, {\bf 1}
& = {}-\pi_3 \, {\bm h} \, {\bm m}_0 \, {\bm h} \, {\bf U} \, {\bf 1}
-\pi_3 \, {\bm h} \, {\bm m}_2 \, {\bm h} \, {\bf U} \, {\bm h} \, {\bf U} \, {\bf 1} \,, \\
{}-\pi_3 \, {\bm h} \, {\bf U} \, {\bm h} \, {\bm m} \,
\frac{1}{{\bf I} -{\bm h} \, {\bf U}} \, {\bf 1}
& = {}-\pi_3 \, {\bm h} \, {\bf U} \, {\bm h} \, {\bm m}_0 \, {\bf 1}
-\pi_3 \, {\bm h} \, {\bf U} \, {\bm h} \, {\bm m}_2 \, {\bm h} \, {\bf U} \, {\bf 1} \,.
\end{split}
\end{equation}
We then expand ${\bm m}_2$ and ${\bm m}_0$ in $g$ to obtain
\begin{equation}
\begin{split}
\pi_3 \, {\bm f} \, {\bf 1}
& = {}-\pi_3 \, {\bm h} \, {\bm m}_0^{(1)} \, {\bm h} \, {\bf U} \, {\bf 1}
-\pi_3 \, {\bm h} \, {\bm b}_2 \, {\bm h} \, {\bf U} \, {\bm h} \, {\bf U} \, {\bf 1} \\
& \quad~ {}-\pi_3 \, {\bm h} \, {\bf U} \, {\bm h} \, {\bm m}_0^{(1)} \, {\bf 1}
-\pi_3 \, {\bm h} \, {\bf U} \, {\bm h} \, {\bm b}_2 \, {\bm h} \, {\bf U} \, {\bf 1}
+O(g^2) \,.
\end{split}
\end{equation}
Among four terms of $O(g)$,
there are two terms which do not involve counterterms.
One of them is ${}-\pi_3 \, {\bm h} \, {\bm b}_2 \, {\bm h} \, {\bf U} \,
{\bm h} \, {\bf U} \, {\bf 1}$:
\begin{equation}
\begin{split}
{}-\pi_3 \, {\bm h} \, {\bm b}_2 \, {\bm h} \, {\bf U} \, {\bm h} \, {\bf U} \, {\bf 1}
& = {}-e^\beta \otimes e^\alpha \otimes h \, b_2 \, ( \, h \, e_\alpha \otimes h \, e_\beta \, ) \\
& \quad~ {}-e^\alpha \otimes e^\beta \otimes h \, b_2 \, ( \, h \, e_\alpha \otimes h \, e_\beta \, ) \\
& \quad~ {}-e^\alpha \otimes h \, e_\alpha \otimes h \, b_2 \, ( \, e^\beta \otimes h \, e_\beta \, ) \\
& = {}-e^\beta \otimes e^\alpha \otimes h \, b_2 \, ( \, h \, e_\alpha \otimes h \, e_\beta \, )
-e^\alpha \otimes e^\beta \otimes h \, b_2 \, ( \, h \, e_\alpha \otimes h \, e_\beta \, ) \\
& \quad~ {}-e^\alpha \otimes h \, e_\alpha \otimes h \, n_0^{(1)} \, {\bf 1} \,.
\end{split}
\end{equation}
The other one is ${}-\pi_3 \, {\bm h} \, {\bf U} \, {\bm h} \, {\bm b}_2 \,
{\bm h} \, {\bf U} \, {\bf 1}$:
\begin{figure}[tb]
\begin{center}
\includegraphics[scale=0.3]{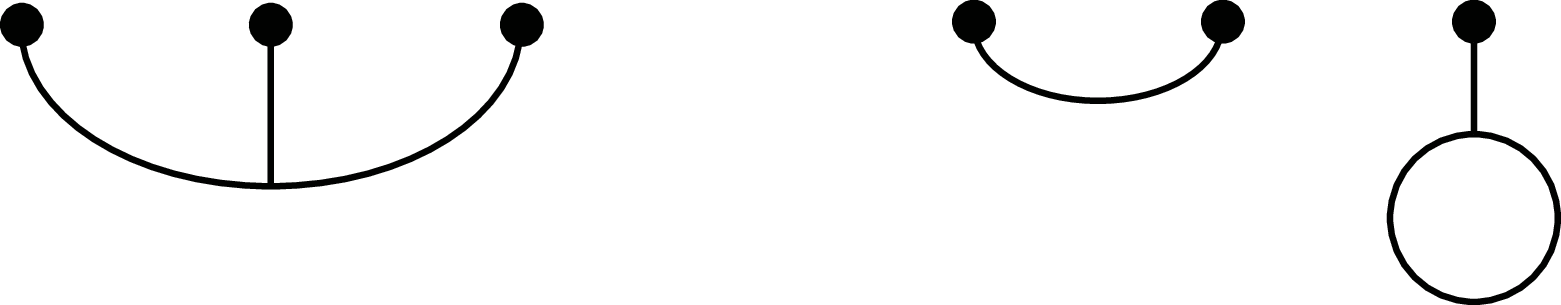}
\end{center}
\caption{Feynman diagrams contributing to the three-point function
$\langle \, \varphi (x_1) \, \varphi (x_2) \, \varphi (x_3) \, \rangle$ at $O(g)$
without involving counterterms.
The left diagram corresponds to the connected part at the tree level
$\langle \, \varphi (x_1) \, \varphi (x_2) \, \varphi (x_3) \, \rangle_C^{(0)}$,
and the right diagram represents a factorized contribution
such as $\langle \, \varphi (x_1) \, \varphi (x_2) \, \rangle^{(0)}
\langle \, \varphi (x_3) \, \rangle^{(1)}$.
}
\label{figure-3-point}
\end{figure}
\begin{equation}
\begin{split}
{}-\pi_3 \, {\bm h} \, {\bf U} \, {\bm h} \, {\bm b}_2 \, {\bm h} \, {\bf U} \, {\bf 1}
& = {}-e^\beta \otimes h \, b_2 \, ( \, e^\alpha \otimes h \, e_\alpha \, ) \otimes h \, e_\beta
-h \, b_2 \, ( \, e^\alpha \otimes h \, e_\alpha \, ) \otimes e^\beta \otimes h \, e_\beta \\
& = {}-e^\beta \otimes h \, n_0^{(1)} \, {\bf 1} \otimes h \, e_\beta
-h \, n_0^{(1)} \, {\bf 1} \otimes e^\beta \otimes h \, e_\beta \,.
\end{split}
\end{equation}
The sum of the two terms is given by
\begin{equation}
\begin{split}
& {}-\pi_3 \, {\bm h} \, {\bm b}_2 \, {\bm h} \, {\bf U} \, {\bm h} \, {\bf U} \, {\bf 1}
{}-\pi_3 \, {\bm h} \, {\bf U} \, {\bm h} \, {\bm b}_2 \, {\bm h} \, {\bf U} \, {\bf 1} \\
& = {}-e^\beta \otimes e^\alpha \otimes h \, b_2 \, ( \, h \, e_\alpha \otimes h \, e_\beta \, )
-e^\alpha \otimes e^\beta \otimes h \, b_2 \, ( \, h \, e_\alpha \otimes h \, e_\beta \, ) \\
& \quad~ {}-e^\alpha \otimes h \, e_\alpha \otimes h \, n_0^{(1)} \, {\bf 1}
{}-e^\alpha \otimes h \, n_0^{(1)} \, {\bf 1} \otimes h \, e_\alpha
-h \, n_0^{(1)} \, {\bf 1} \otimes e^\alpha \otimes h \, e_\alpha \,.
\end{split}
\end{equation}
The remaining two terms involving counterterms are given by
\begin{align}
{}-\pi_3 \, {\bm h} \, {\bm m}_0^{(1)} \, {\bm h} \, {\bf U} \, {\bf 1}
& = {}-e^\alpha \otimes h \, e_\alpha \otimes h \, m_0^{(1)} \, {\bf 1} \,, \\
{}-\pi_3 \, {\bm h} \, {\bf U} \, {\bm h} \, {\bm m}_0^{(1)} \, {\bf 1}
& = {}-e^\alpha \otimes h \, m_0^{(1)} \, {\bf 1} \otimes h \, e_\alpha
-h \, m_0^{(1)} \, {\bf 1} \otimes e^\alpha \otimes h \, e_\alpha \,.
\end{align}
We therefore find
\begin{equation}
\begin{split}
\pi_3 \, {\bm f} \, {\bf 1}
& = {}-e^\beta \otimes e^\alpha \otimes h \, b_2 \, ( \, h \, e_\alpha \otimes h \, e_\beta \, )
-e^\alpha \otimes e^\beta \otimes h \, b_2 \, ( \, h \, e_\alpha \otimes h \, e_\beta \, ) \\
& \quad~ {}-e^\alpha \otimes h \, e_\alpha \otimes h \, \Gamma_0^{(1)} \, {\bf 1}
-e^\alpha \otimes h \, \Gamma_0^{(1)} \, {\bf 1} \otimes h \, e_\alpha
-h \, \Gamma_0^{(1)} \, {\bf 1} \otimes e^\alpha \otimes h \, e_\alpha
+O(g^2) \,.
\end{split}
\end{equation}
The three-point function is given by
\begin{equation}
\begin{split}
& \langle \, \varphi (x_1) \, \varphi (x_2) \, \varphi (x_3) \, \rangle
= \omega_3 \, ( \, \pi_3 \, {\bm f} \, {\bf 1} \,,
\delta^d (x-x_1) \otimes \delta^d (x-x_2) \otimes \delta^d (x-x_3) \, ) \\
& = \langle \, \varphi (x_1) \, \varphi (x_2) \, \varphi (x_3) \, \rangle_C^{(0)}
+\langle \, \varphi (x_1) \, \varphi (x_2) \, \rangle^{(0)}
\langle \, \varphi (x_3) \, \rangle^{(1)} \\
& \quad~ +\langle \, \varphi (x_1) \, \varphi (x_3) \, \rangle^{(0)}
\langle \, \varphi (x_2) \, \rangle^{(1)}
+\langle \, \varphi (x_2) \, \varphi (x_3) \, \rangle^{(0)}
\langle \, \varphi (x_1) \, \rangle^{(1)}
+O(g^2) \,,
\end{split}
\end{equation}
where the connected part at the tree level
$\langle \, \varphi (x_1) \, \varphi (x_2) \, \varphi (x_3) \, \rangle_C^{(0)}$
is given by
\begin{equation}
\begin{split}
& \langle \, \varphi (x_1) \, \varphi (x_2) \, \varphi (x_3) \, \rangle_C^{(0)} \\
& = g \int \frac{d^d p_1}{(2 \pi)^d} \int \frac{d^d p_2}{(2 \pi)^d} \,
\frac{e^{-ip_1 x_1 -ip_2 x_2 +i (p_1+p_2) \, x_3}}
{( \, p_1^2 +m^2 \, ) \, ( \, p_2^2 +m^2 \, ) \, ( \, ( \, p_1+p_2 \, )^2 +m^2 \, )} \,.
\end{split}
\end{equation}
See figure~\ref{figure-3-point}.

\section{The Schwinger-Dyson equations}
\label{section-5}
\setcounter{equation}{0}

We have demonstrated that our formula
presented in subsection~\ref{subsection-2.4} correctly reproduces
correlation functions of $\varphi^3$ theory.
In this section we show that the Schwinger-Dyson equations are satisfied.

In the framework of the path integral, correlation functions are defined by
\begin{equation}
\langle \, \varphi (x_1 ) \, \varphi (x_2) \, \ldots \, \varphi (x_n) \, \rangle
= \frac{1}{Z} \int \mathcal{D} \varphi \,
\varphi (x_1 ) \, \varphi (x_2) \, \ldots \, \varphi (x_n) \, e^{-S} \,,
\label{Euclidean-correlation-functions}
\end{equation}
where
\begin{equation}
Z = \int \mathcal{D} \varphi \, e^{-S} \,.
\end{equation}
Since
\begin{equation}
\frac{1}{Z} \int \mathcal{D} \varphi \, \frac{\delta}{\delta \varphi (y)} \, \biggl[ \,
\varphi (x_1 ) \, \varphi (x_2) \, \ldots \, \varphi (x_n) \, e^{-S} \, \biggr] = 0 \,,
\end{equation}
we obtain the Schwinger-Dyson equations given by
\begin{equation}
\sum_{i=1}^n \delta^d ( y-x_i ) \, \langle \, \varphi (x_1) \, \ldots \,
\varphi (x_{i-1}) \, \varphi (x_{i+1}) \, \ldots \, \varphi (x_n ) \, \rangle 
-\langle \, \varphi (x_1) \, \ldots \,
\varphi (x_n ) \, \frac{\delta S}{\delta \varphi (y)} \, \rangle = 0 \,.
\end{equation}

Let us show that correlation functions described in terms of quantum $A_\infty$ algebras
satisfy the Schwinger-Dyson equations.
Since
\begin{equation}
( \, {\bf I} +{\bm h} \, {\bm m} -{\bm h} \, {\bf U} \, ) \,
\frac{1}{{\bf I} +{\bm h} \, {\bm m} -{\bm h} \, {\bf U}} \, {\bf 1} = {\bf 1}
\label{structure-Euclidean}
\end{equation}
and
\begin{equation}
\pi_{n+1} \, {\bf 1} = 0 \quad \text{for} \quad n \ge 0 \,,
\end{equation}
we have
\begin{equation}
\pi_{n+1} \, {\bm f} \, {\bf 1}
+\pi_{n+1} \, {\bm h} \, {\bm m} \, {\bm f} \, {\bf 1}
-\pi_{n+1} \, {\bm h} \, {\bf U} \, {\bm f} \, {\bf 1} = 0 \quad
\text{for} \quad n \ge 0 \,.
\label{SD}
\end{equation}
The first term on the left-hand side gives the following $(n+1)$-point function:
\begin{equation}
\begin{split}
& \omega_{n+1} \, ( \, \pi_{n+1} \, {\bm f} \, {\bf 1} \,,
\delta^d (x-x_1) \otimes \delta^d (x-x_2) \otimes \ldots
\otimes \delta^d (x-x_n) \otimes \delta^d (x-y) \, ) \\
& = \langle \, \varphi (x_1) \, \varphi (x_2) \, \ldots \, \varphi (x_n) \, \varphi (y) \, \rangle \,,
\end{split}
\label{SD-1}
\end{equation}
where we have chosen the last argument to be $y$ instead of $x_{n+1}$.
Let us next consider
\begin{equation}
\omega_{n+1} \, ( \, \pi_{n+1} \, {\bm h} \, {\bf U} \, {\bm f} \, {\bf 1} \,,
\delta^d (x-x_1) \otimes \delta^d (x-x_2) \otimes \ldots
\otimes \delta^d (x-x_n) \otimes \delta^d (x-y) \, )
\end{equation}
from the third term
on the left-hand side of~\eqref{SD}.
It follows from the decomposition of ${\bm h} \, {\bf U}$ in~\eqref{hU-decomposition} that
\begin{equation}
\begin{split}
& \omega_{n+1} \, ( \, \pi_{n+1} \, {\bm h} \, {\bf U} \, {\bm f} \, {\bf 1} \,,
\delta^d (x-x_1) \otimes \delta^d (x-x_2) \otimes \ldots
\otimes \delta^d (x-x_n) \otimes \delta^d (x-y) \, ) \\
& = \omega_{n+1} \, ( \, {\bm h} \, {\bf U} \, \pi_{n-1} \, {\bm f} \, {\bf 1} \,,
\delta^d (x-x_1) \otimes \delta^d (x-x_2) \otimes \ldots
\otimes \delta^d (x-x_n) \otimes \delta^d (x-y) \, )
\end{split}
\end{equation}
for $n \ge 1$.
The action of ${\bm h} \, {\bf U}$ in~\eqref{hU-action} gives factors of the form
\begin{equation}
\int dx'_i \int dy' \int \frac{d^d p}{(2 \pi)^d} \,
e^{-ip x'_i} \, \frac{1}{p^2+m^2} \, e^{ip y'} \, \delta^d (x'_i-x_i) \, \delta^d (y'-y)
= \int \frac{d^d p}{(2 \pi)^d} \,
\frac{e^{-ip \, (x_i-y)}}{p^2+m^2}
\end{equation}
so that we find
\begin{equation}
\begin{split}
& \omega_{n+1} \, ( \, \pi_{n+1} \, {\bm h} \, {\bf U} \, {\bm f} \, {\bf 1} \,,
\delta^d (x-x_1) \otimes \delta^d (x-x_2) \otimes \ldots
\otimes \delta^d (x-x_n) \otimes \delta^d (x-y) \, ) \\
& = \sum_{i=1}^n \int \frac{d^d p}{(2 \pi)^d} \,
\frac{e^{-ip \, (x_i-y)}}{p^2+m^2} \, \langle \, \varphi (x_1) \, \ldots \,
\varphi (x_{i-1}) \, \varphi (x_{i+1}) \, \ldots \, \varphi (x_n ) \, \rangle \,.
\end{split}
\label{SD-3}
\end{equation}
The second term on the left-hand side of~\eqref{SD} is expanded as
\begin{equation}
\pi_{n+1} \, {\bm h} \, {\bm m} \, {\bm f} \, {\bf 1}
= \sum_{k=0}^\infty \pi_{n+1} \, {\bm h} \, {\bm m}_k \, {\bm f} \, {\bf 1}
= \sum_{k=0}^\infty \,
( \, {\mathbb I}^{\otimes n} \otimes h \, m_k \, ) \, \pi_{n+k} \, {\bm f} \, {\bf 1} \,.
\end{equation}
If we compare
\begin{equation}
\omega_{n+1} \, ( \, ( \, {\mathbb I}^{\otimes n} \otimes h \, m_k \, ) \, \pi_{n+k} \, {\bm f} \, {\bf 1} \,,
\delta^d (x-x_1) \otimes \delta^d (x-x_2) \otimes \ldots
\otimes \delta^d (x-x_n) \otimes \delta^d (x-y) \, )
\label{hm_k}
\end{equation}
with
\begin{equation}
\begin{split}
& \omega_{n+k} \, ( \, \pi_{n+k} \, {\bm f} \, {\bf 1} \,,
\delta^d (x-x_1) \otimes \delta^d (x-x_2) \otimes \ldots
\otimes \delta^d (x-x_{n+k}) \, ) \\
& = \langle \, \varphi (x_1) \, \varphi (x_2) \, \ldots \, \varphi (x_{n+k}) \, \rangle \,,
\end{split}
\label{n+k}
\end{equation}
the action of $h \, m_k$ in~\eqref{hm_k} corresponds
to replacing $\varphi (x_{n+1}) \, \varphi (x_{n+2}) \, \ldots \, \varphi (x_{n+k})$
{\it inside} the brackets of the correlation function~\eqref{n+k} with
\begin{equation}
\int d^d z \int \frac{d^d p}{(2 \pi)^d} \, \frac{e^{-ip \, (y-z)}}{p^2+m^2} \,
m_k \, ( \, \varphi (z) \otimes \ldots \, \otimes  \varphi (z) \, )
\end{equation}
for $k \ge 1$, where $m_k \, ( \, \varphi (z) \otimes \ldots \, \otimes  \varphi (z) \, )$
represents the function of $z$ obtained by replacing
$x$ in $m_k \, ( \, \varphi (x) \otimes \ldots \, \otimes  \varphi (x) \, )$ with $z$.
The action of $h \, m_0$ in~\eqref{hm_k} corresponds to multiplying
$\langle \, \varphi (x_1) \, \varphi (x_2) \, \ldots \, \varphi (x_n) \, \rangle$ by
\begin{equation}
\int d^d z \int \frac{d^d p}{(2 \pi)^d} \, \frac{e^{-ip \, (y-z)}}{p^2+m^2} \,
m_0 \, {\bf 1} \Bigr|_{x=z} \,.
\end{equation}
In what follows the notation
$m_k \, ( \, \varphi (z) \otimes \ldots \, \otimes  \varphi (z) \, )$
with $k=0$
represents the function of $z$ obtained by replacing $x$ in ${\bm m}_0 \, {\bf 1}$ with $z$.
We thus conclude that
\begin{equation}
\begin{split}
& \omega_{n+1} \, ( \, \pi_{n+1} \, {\bm h} \, {\bm m}_k \, {\bm f} \, {\bf 1} \,,
\delta^d (x-x_1) \otimes \delta^d (x-x_2) \otimes \ldots
\otimes \delta^d (x-x_n) \otimes \delta^d (x-y) \, ) \\
& = \int d^d z \int \frac{d^d p}{(2 \pi)^d} \, \frac{e^{-ip \, (y-z)}}{p^2+m^2} \,
\langle \, \varphi (x_1) \, \varphi (x_2) \, \ldots \, \varphi (x_n) \, 
m_k \, ( \, \varphi (z) \otimes \ldots \, \otimes  \varphi (z) \, ) \, \rangle \,.
\end{split}
\label{SD-2}
\end{equation}
Using~\eqref{SD-1}, \eqref{SD-3}, and~\eqref{SD-2},
the relation~\eqref{SD} implies that
\begin{equation}
\begin{split}
& \langle \, \varphi (x_1) \ldots \, \varphi (x_n) \, \varphi (y) \, \rangle \\
& +\sum_{k=0}^\infty
\int d^d z \int \frac{d^d p}{(2 \pi)^d} \, \frac{e^{-ip \, (y-z)}}{p^2+m^2} \,
\langle \, \varphi (x_1) \ldots \, \varphi (x_n) \,
m_k \, ( \, \varphi (z) \otimes \ldots \, \otimes  \varphi (z) \, ) \, \rangle \\
& {}-\sum_{i=1}^n \int \frac{d^d p}{(2 \pi)^d} \,
\frac{e^{-ip \, (x_i-y)}}{p^2+m^2} \, \langle \, \varphi (x_1) \, \ldots \,
\varphi (x_{i-1}) \, \varphi (x_{i+1}) \, \ldots \, \varphi (x_n ) \, \rangle = 0 \,.
\end{split}
\end{equation}
We then acts the operator ${}-\partial_y^2 +m^2$ to find
\begin{equation}
\begin{split}
& ( \, {}-\partial_y^2 +m^2 \, ) \,
\langle \, \varphi (x_1) \ldots \, \varphi (x_n) \, \varphi (y) \, \rangle \\
& +\sum_{k=0}^\infty \langle \, \varphi (x_1) \ldots \, \varphi (x_n) \,
m_k \, ( \, \varphi (y) \otimes \ldots \, \otimes  \varphi (y) \, ) \, \rangle \\
& {}-\sum_{i=1}^n \delta^d (y-x_i) \,
\langle \, \varphi (x_1) \, \ldots \,
\varphi (x_{i-1}) \, \varphi (x_{i+1}) \, \ldots \, \varphi (x_n ) \, \rangle = 0 \,.
\end{split}
\end{equation}
Since
\begin{equation}
\frac{\delta S}{\delta \varphi (y)}
= ( \, {}-\partial_y^2 +m^2 \, ) \, \varphi (y)
+\sum_{k=0}^\infty m_k \, ( \, \varphi (y) \otimes \ldots \, \otimes  \varphi (y) \, ) \,,
\end{equation}
we find
\begin{equation}
{}-\sum_{i=1}^n \delta^d ( y-x_i ) \, \langle \, \varphi (x_1) \, \ldots \,
\varphi (x_{i-1}) \, \varphi (x_{i+1}) \, \ldots \, \varphi (x_n ) \, \rangle 
+\langle \, \varphi (x_1) \, \ldots \,
\varphi (x_n ) \, \frac{\delta S}{\delta \varphi (y)} \, \rangle = 0 \,.
\end{equation}
We have thus shown that the Schwinger-Dyson equations are satisfied.

\section{Scalar field theories in Minkowski space}
\label{section-6}
\setcounter{equation}{0}

So far we have considered scalar field theories in Euclidean space.
In this section we consider scalar field theories in Minkowski space.
The action of the free theory is given by
\begin{equation}
{}-\frac{1}{2} \int d^d x \, [ \, \partial_\mu \varphi (x) \,  \partial^\mu \varphi (x)
+m^2 \, \varphi (x)^2 \, ] \,.
\end{equation}
This can be written as
\begin{equation}
{}-\frac{1}{2} \int d^d x \, [ \, \partial_\mu \varphi (x) \,  \partial^\mu \varphi (x)
+m^2 \, \varphi (x)^2 \, ]
= {}-\frac{1}{2} \, \omega \, ( \, \varphi (x), Q \, \varphi (x) \, )
\end{equation}
with the understanding that $\partial^2$ in the definition of $Q$
is changed from $g_{\mu \nu} \, \partial^\mu \partial^\nu$
with the Euclidean metric $g_{\mu \nu}$
to $\eta_{\mu \nu} \, \partial^\mu \partial^\nu$
with the Minkowski metric $\eta_{\mu \nu}$ of signature $(-,+,\cdots,+) \,$.
The equation of motion of the free theory is given by
\begin{equation}
Q \, \varphi (x) = ( \, {}-\partial^2 +m^2 \, ) \, \varphi (x) = 0
\end{equation}
for $\varphi (x)$ in $\mathcal{H}_1$.
Unlike the Euclidean case where the solution is unique,
solutions to the equation of motion
consist of general superpositions of propagating waves.
When we consider correlation functions, however, we claim that we should consider
the projection onto $\varphi (x) = 0$
so that the corresponding projection operator $P$ vanishes:
\begin{equation}
P = 0 \,.
\end{equation}
As we wrote in subsection~\ref{subsection-2.4},
this should correspond to carrying out the path integral completely,
and this should also be the case for the theory in Minkowski space.
The conditions for $h$ are again given by
\begin{equation}
Q \, h +h \, Q = {\mathbb I} \,, \qquad
h^2 = 0 \,.
\end{equation}
To define the path integral of the free theory in Minkowski space,
we use the $i \epsilon$ prescription
and as a result we obtain the Feynman propagator.
Since we define correlation functions in Minkowski space
as vacuum expectation values associated with the unique vacuum in the quantum theory,
we use the Feynman propagation to define the operator $h$.
The action of $h$ on $\varphi (x)$ in $\mathcal{H}_2$ is given by
\begin{equation}
h \, \varphi (x) = \int d^d y \int \frac{d^d p}{(2 \pi)^d} \,
\frac{e^{-ip \, (x-y)}}{p^2+m^2-i \epsilon} \, \varphi (y) \,,
\end{equation}
and the operator $h$ annihilates any element in $\mathcal{H}_1$.

We consider an action of the form
\begin{equation}
S = {}-\frac{1}{2} \, \omega \, ( \, \varphi (x), Q \, \varphi (x) \, )
-\sum_{n=0}^\infty \, \frac{1}{n+1} \,
\omega \, ( \, \varphi (x) \,, m_n \, ( \, \varphi (x) \otimes \ldots \otimes \varphi (x) \, ) \, ) \,. 
\end{equation}
Note that the overall sign has been changed from the Euclidean case.
We claim that correlation functions are given by
\begin{equation}
\langle \, \varphi (x_1) \, \varphi (x_2) \, \ldots \, \varphi (x_n) \, \rangle
= \omega_n \, ( \, \pi_n \, {\bm f} \, {\bf 1} \,,
\delta^d (x-x_1) \otimes \delta^d (x-x_2) \otimes \ldots \otimes \delta^d (x-x_n) \, ) \,,
\end{equation}
where
\begin{equation}
{\bm f} = \frac{1}{{\bf I} +{\bm h} \, {\bm m} +i \hbar \, {\bm h} \, {\bf U}} \,.
\end{equation}

Let us show that the Schwinger-Dyson equations are satisfied.
In the framework of the path integral, correlation functions are defined by
\begin{equation}
\langle \, \varphi (x_1 ) \, \varphi (x_2) \, \ldots \, \varphi (x_n) \, \rangle
= \frac{1}{Z} \int \mathcal{D} \varphi \,
\varphi (x_1 ) \, \varphi (x_2) \, \ldots \, \varphi (x_n) \, e^{\, \frac{i}{\hbar} S} \,,
\label{Minkowski-correlation-functions}
\end{equation}
where
\begin{equation}
Z = \int \mathcal{D} \varphi \, e^{\, \frac{i}{\hbar} S} \,.
\end{equation}
Since
\begin{equation}
\frac{1}{Z} \int \mathcal{D} \varphi \, \frac{\delta}{\delta \varphi (y)} \, \biggl[ \,
\varphi (x_1 ) \, \varphi (x_2) \, \ldots \, \varphi (x_n) \, e^{\, \frac{i}{\hbar} S} \, \biggr] = 0 \,,
\end{equation}
we obtain the Schwinger-Dyson equations given by
\begin{equation}
\sum_{i=1}^n \delta^d ( y-x_i ) \, \langle \, \varphi (x_1) \, \ldots \,
\varphi (x_{i-1}) \, \varphi (x_{i+1}) \, \ldots \, \varphi (x_n ) \, \rangle 
+\frac{i}{\hbar} \, \langle \, \varphi (x_1) \, \ldots \,
\varphi (x_n ) \, \frac{\delta S}{\delta \varphi (y)} \, \rangle = 0 \,.
\end{equation}

Let us show that correlation functions described in terms of quantum $A_\infty$ algebras
satisfy the Schwinger-Dyson equations.
Since
\begin{equation}
( \, {\bf I} +{\bm h} \, {\bm m} +i \hbar \, {\bm h} \, {\bf U} \, ) \,
\frac{1}{{\bf I} +{\bm h} \, {\bm m} +i \hbar \, {\bm h} \, {\bf U}} \, {\bf 1} = {\bf 1}
\label{structure-Minkowski}
\end{equation}
and
\begin{equation}
\pi_{n+1} \, {\bf 1} = 0 \quad \text{for} \quad n \ge 0 \,,
\end{equation}
we have
\begin{equation}
\begin{split}
& \pi_{n+1} \, {\bm f} \, {\bf 1}
+\pi_{n+1} \, {\bm h} \, {\bm m} \, {\bm f} \, {\bf 1}
+i \hbar \, \pi_{n+1} \, {\bm h} \, {\bf U} \, {\bm f} \, {\bf 1} \\
& = \pi_{n+1} \, {\bm f} \, {\bf 1}
+\sum_{k=0}^\infty \pi_{n+1} \, {\bm h} \, {\bm m}_k \, {\bm f} \, {\bf 1}
+i \hbar \, \pi_{n+1} \, {\bm h} \, {\bf U} \, {\bm f} \, {\bf 1} = 0 \quad
\text{for} \quad n \ge 0 \,.
\end{split}
\end{equation}
Using~\eqref{SD-1}, \eqref{SD-3}, and~\eqref{SD-2}
with $p^2 +m^2$ replaced by $p^2 +m^2 -i \epsilon$,
we obtain
\begin{equation}
\begin{split}
& \langle \, \varphi (x_1) \ldots \, \varphi (x_n) \, \varphi (y) \, \rangle \\
& +\sum_{k=0}^\infty
\int d^d z \int \frac{d^d p}{(2 \pi)^d} \, \frac{e^{-ip (y-z)}}{p^2+m^2 -i \epsilon} \,
\langle \, \varphi (x_1) \ldots \, \varphi (x_n) \,
m_k \, ( \, \varphi (z) \otimes \ldots \, \otimes  \varphi (z) \, ) \, \rangle \\
& +i \hbar \sum_{i=1}^n \int \frac{d^d p}{(2 \pi)^d} \,
\frac{e^{-ip \, (x_i-y)}}{p^2+m^2 -i \epsilon} \, \langle \, \varphi (x_1) \, \ldots \,
\varphi (x_{i-1}) \, \varphi (x_{i+1}) \, \ldots \, \varphi (x_n ) \, \rangle = 0 \,.
\end{split}
\end{equation}
We then acts the operator ${}-\partial_y^2 +m^2$ to find
\begin{equation}
\begin{split}
& ( \, {}-\partial_y^2 +m^2 \, ) \,
\langle \, \varphi (x_1) \ldots \, \varphi (x_n) \, \varphi (y) \, \rangle \\
& +\sum_{k=0}^\infty \langle \, \varphi (x_1) \ldots \, \varphi (x_n) \,
m_k \, ( \, \varphi (y) \otimes \ldots \, \otimes  \varphi (y) \, ) \, \rangle \\
& +i \hbar \sum_{i=1}^n \delta^d (y-x_i) \,
\langle \, \varphi (x_1) \, \ldots \,
\varphi (x_{i-1}) \, \varphi (x_{i+1}) \, \ldots \, \varphi (x_n ) \, \rangle = 0 \,.
\end{split}
\end{equation}
Since
\begin{equation}
\frac{\delta S}{\delta \varphi (y)}
= {}-( \, {}-\partial_y^2 +m^2 \, ) \, \varphi (y)
-\sum_{k=0}^\infty m_k \, ( \, \varphi (y) \otimes \ldots \, \otimes  \varphi (y) \, ) \,,
\end{equation}
we find
\begin{equation}
i \hbar \sum_{i=1}^n \delta^d ( y-x_i ) \, \langle \, \varphi (x_1) \, \ldots \,
\varphi (x_{i-1}) \, \varphi (x_{i+1}) \, \ldots \, \varphi (x_n ) \, \rangle 
-\langle \, \varphi (x_1) \, \ldots \,
\varphi (x_n ) \, \frac{\delta S}{\delta \varphi (y)} \, \rangle = 0 \,.
\end{equation}
We have thus shown that the Schwinger-Dyson equations are satisfied.

Let us consider the two-point function of the free theory.
In this case the coderivation ${\bm m}$ vanishes
and ${\bm f}$ is given by
\begin{equation}
{\bm f} = \frac{1}{{\bf I} +i \hbar \, {\bm h} \, {\bf U}} \,.
\end{equation}
The two-point function can be calculated from $\pi_2 \, {\bm f} \, {\bf 1}$. We find
\begin{equation}
\pi_2 \, {\bm f} \, {\bf 1}
= {}-i \hbar \, \pi_2 \, {\bm h} \, {\bf U} \, {\bf 1}
= {}-i \hbar \, e^\alpha \otimes h \, e_\alpha
= {}-i \hbar \, \int \frac{d^d p}{(2 \pi)^d} \,
e^{-ipx} \otimes \frac{1}{p^2+m^2 -i \epsilon} \, e^{ipx} \,.
\end{equation}
The two-point function is then given by
\begin{equation}
\begin{split}
& \langle \, \varphi (x_1) \, \varphi (x_2) \, \rangle
= \omega_2 \, ( \, \pi_2 \, {\bm f} \, {\bf 1} \,,
\delta^d (x-x_1) \otimes \delta^d (x-x_2) \, ) \\
& = {}-i \hbar \int \frac{d^d p}{(2 \pi)^d} \,
\frac{e^{-ip \, (x_1-x_2)}}{p^2+m^2 -i \epsilon} \,.
\end{split}
\end{equation}
More examples of calculations for scalar field theories in Minkowski space
will be presented in~\cite{Okawa-Shibuya}.
 
Quantum mechanics corresponds to the case where $d=1$.
We write the action of a harmonic oscillator as
\begin{equation}
S = \int dt \, \biggl[ \, \frac{1}{2} \, m \, \biggl( \, \frac{d q(t)}{dt} \, \biggr)^2
-\frac{1}{2} \, m \, \omega^2 \, q(t)^2 \, \biggr] \,,
\end{equation}
where the parameters $m$ and $\omega$ are real and positive.
We take $\mathcal{H}_1$ and $\mathcal{H}_2$ to be the vector space of functions of $t$,
and we define $Q$ by
\begin{equation}
Q \, q(t) = m \, \biggl( \, \frac{d^2}{dt^2} +\omega^2 \, \biggr) \, q(t)
\end{equation}
for $q(t)$ in $\mathcal{H}_1$.
We define $h$ by
\begin{equation}
h \, q(t) = \frac{1}{m} \int dt' \int \frac{d \omega'}{2 \pi} \,
\frac{e^{i \omega' (t-t')}}{{}-\omega'^{\, 2} +\omega^2 -i \epsilon} \, q(t')
\end{equation}
for $q(t)$ in $\mathcal{H}_2$.
For quantum mechanics we use the following choice for $e^\alpha$ and $e_\alpha$
which appear in $T \mathcal{H}$:
\begin{equation}
\begin{split}
& \ldots \otimes e^\alpha \otimes \ldots \otimes e_\alpha \otimes \ldots
= \int \frac{d \omega'}{2 \pi} \,
\ldots \otimes e^{i \omega' t} \otimes \ldots \otimes e^{-i \omega' t} \otimes \ldots \,.
\end{split}
\end{equation}
The two-point function is given by
\begin{equation}
\langle \, q(t_1) \, q(t_2) \, \rangle
= {}-\frac{i \hbar}{m} \int \frac{d \omega'}{2 \pi} \,
\frac{e^{i \omega' (t_1-t_2)}}{{}-\omega'^{\, 2} +\omega^2 -i \epsilon} \,,
\end{equation}
which is
\begin{equation}
\langle \, q(t_1) \, q(t_2) \, \rangle
= \frac{\hbar}{2 \, m \omega} \, e^{-i \omega \, (t_1-t_2)} \quad \text{for} \quad t_1 > t_2
\end{equation}
and
\begin{equation}
\langle \, q(t_1) \, q(t_2) \, \rangle
= \frac{\hbar}{2 \, m \omega} \, e^{i \omega \, (t_1-t_2)} \quad \text{for} \quad t_1 < t_2 \,.
\end{equation}
As is well known, this reproduces the vacuum expectation value of
the time-ordered product
$\langle 0 | \, T \, \hat{q} (t_1) \, \hat{q} (t_2) \, | 0 \rangle$ 
for
\begin{equation}
\hat{q} (t) = \sqrt{\frac{\hbar}{2 \, m \omega}} \,
( \, \hat{a} \, e^{-i \omega t} +\hat{a}^\dagger \, e^{i \omega t} \, )
\end{equation}
with
\begin{equation}
[ \, \hat{a}, \hat{a}^\dagger \, ] = 1 \,, \qquad \hat{a} \, | 0 \rangle = 0 \,.
\end{equation}

\section{Conclusions and discussion}
\label{section-7}
\setcounter{equation}{0}

In this paper we proposed the formula~\eqref{formula}
for correlation function of scalar field theories in perturbation theory
using quantum $A_\infty$ algebras.
We then proved that correlation functions from our formula
satisfy the Schwinger-Dyson equations
as an immediate consequence of the structure in~\eqref{structure-Euclidean}
for the Euclidean case
and in~\eqref{structure-Minkowski}
for the Minkowski case.

Since the description in terms of homotopy algebras or the Batalin-Vilkovisky formalism
tends to be elusive and formal,
we have presented completely explicit calculations for $\varphi^3$ theory
which involve renormalization at one loop.
We hope that this demonstration in this paper helps us
convince ourselves that any calculations of this kind
in the path integral or in the operator formalism
can be carried out in the framework of quantum $A_\infty$ algebras as well.

The important ingredient ${\bm f}$ is associated with a quasi-isomorphism from the $A_\infty$ algebra
after the projection to the $A_\infty$ algebra before the projection.
While $\pi_1 \, {\bm f}$ describes the quasi-isomorphism
and we are usually interested in this part of ${\bm f}$,
we found that the part ${\bm f} \, \pi_0$ is relevant for correlation functions.
Incidentally, the sector $\mathcal{H}^{\otimes 0}$ is often omitted
in the discussion of homotopy algebras, but it plays an important role in our approach.
For any $A_\infty$ algebra described in terms of a coderivation ${\bf Q}+{\bm m}$,
the minimal model theorem~\cite{Kajiura:2003ax} states
the existence of a quasi-isomorphism from an $A_\infty$ algebra on the cohomology of $Q$
to the $A_\infty$ algebra described in terms of ${\bf Q}+{\bm m}$.
Such an $A_\infty$ algebra on the cohomology of $Q$ is called a minimal model
of the $A_\infty$ algebra described in terms of ${\bf Q}+{\bm m}$,
and the minimal model is known to be unique up to isomorphisms.
While we used the perturbative expression of ${\bm f}$ based on the homological perturbation lemma,
we hope that the characterization in terms of ${\bm f}$
leads to the nonperturbative definition of correlation functions.
In particular, it would be interesting to address the question
of how the definition of the path integral
based on Lefschetz thimbles~\cite{Witten:2010cx}
can be incorporated into the framework of homotopy algebras.

As we mentioned in the introduction,
correlation functions were discussed
in the framework of the Batalin-Vilkovisky formalism~\cite{Gwilliam:2012jg, Chiaffrino:2021pob}.
Quantum $L_\infty$ algebras, discussed for example in~\cite{Doubek:2017naz},
involve symmetrization procedures
and are more naturally related to the Batalin-Vilkovisky formalism.
We chose quantum $A_\infty$ algebras,
and what was surprising was that correlation functions
which are symmetric under the exchange of scalar fields
are obtained without any symmetrization procedures.\footnote{
The correlation functions which we reproduce by our formula are
neither ``color-ordered'' nor amputated,
and they are defined in the path integral formalism
by~\eqref{Euclidean-correlation-functions}
for theories in Euclidean space
and by~\eqref{Minkowski-correlation-functions}
for theories in Minkowski space.
The $n$-point function
$\langle \, \varphi (x_1 ) \, \varphi (x_2) \, \ldots \, \varphi (x_n) \, \rangle$
is totally symmetric with respect to $x_1$, $x_2$, \ldots, and $x_n$.
}
First, the construction of the vector space $T \mathcal{H}$
does not involve symmetrization procedures
unlike the corresponding vector space for $L_\infty$ algebras,
and elements in~$\mathcal{H}^{\otimes n}$ are generically not graded symmetric.
However, the action of ${\bf U}$ symmetrizes the resulting element
when it acts on a symmetrized element so that elements of the form ${\bf U}^n \, {\bf 1}$,
for example, are graded symmetric.
Our formula for correlation functions uses ${\bf U}^n \, {\bf 1}$ as building blocks,
and this is part of the reason why our formula
reproduces symmetric correlation functions,
but our formula also involves ${\bm m}$ and ${\bm h}$
which obscure the symmetric nature at intermediate steps.
As can be seen from the definition~\eqref{c_n-pi_m},
coderivations in $A_\infty$ algebras
are in accord with cyclic properties but do not involve symmetrization procedures,
so we do not expect that the coderivation ${\bm m}$ in our formula
preserves the symmetric property of the elements generated
by actions of~${\bf U}$ from~${\bf 1}$.
Furthermore, the definition of $\bm{h}$ is asymmetric
as we commented below~\eqref{boldface-h}.
Since $P=0$ in our formula, only the last term
on the right-hand side of~\eqref{boldface-h} survives
and the rightmost sector of $\mathcal{H}^{\otimes n}$ plays a distinguished role.
This is reflected in~\eqref{hU-action} for ${\bm h} \, {\bf U}$
and in~\eqref{hm_0-action}, \eqref{hm_1-action}, and~\eqref{hm_2-action}
for ${\bm h} \, {\bm m}_0$, ${\bm h} \, {\bm m}_1$,
and ${\bm h} \, {\bm m}_2$, respectively.
Nevertheless, it turned out that $\pi_n \, {\bm f} \, {\bf 1}$
is totally symmetric and gives correlation functions
which are symmetric under the exchange of scalar fields.
This remarkable property has made our formula simpler,
and it would be technically useful
in the generalization to open string field theory.
Since correlation functions based on our formula
satisfy the Schwinger-Dyson equations,
they must be symmetric under the exchange of scalar fields,
but currently we only have this indirect understanding,
and it would be important to unveil the hidden structure of ${\bm f}$.

While the expressions for correlation functions in terms of homotopy algebras
are universal,
our expressions are restricted to the case
where $\mathcal{H}$ consists of only two sectors $\mathcal{H}_1$ and $\mathcal{H}_2$.
The property that the operator $h$ annihilates any element
in $\mathcal{H}_1$ simplified the calculations, but this is not the case
for general $A_\infty$ algebras.
It would be important to extend our analysis to more general cases.

Our ultimate goal is to provide a framework to prove the AdS/CFT correspondence
using open string field theory with source terms for gauge-invariant operators
following the scenario outlined in~\cite{Okawa:2020llq}.
The quantum treatment of open string field theory must be crucial
for this program, and we hope that quantum $A_\infty$ algebras will provide us
with powerful tools in this endeavor.

\bigskip
\noindent
{\normalfont \bfseries \large Acknowledgments}

\medskip
This work was supported in part by JSPS KAKENHI Grant Number JP17K05408.


\small

\begin{thebibliography}{99}

\bibitem{Stasheff:I}
  J.~D.~Stasheff,
  ``Homotopy associativity of $H$-spaces. I,''
  Trans. of the Amer. Math. Soc. {\bf 108}, 275 (1963).

\bibitem{Stasheff:II}
  J.~D.~Stasheff,   ``Homotopy associativity of $H$-spaces. II,''
  Trans. of the Amer. Math. Soc. {\bf 108}, 293 (1963).

\bibitem{Getzler-Jones}
  E.~Getzler and J.~D.~S.~Jones,
  ``$A_\infty$-algebras and the cyclic bar complex,''
  Illinois~J.~Math {\bf 34}, 256 (1990).

\bibitem{Markl}
  M.~Markl,
  ``A cohomology theory for $A (m)$-algebras and applications,''
  J. Pure Appl. Algebra {\bf 83}, 141 (1992).

\bibitem{Penkava:1994mu}
  M.~Penkava and A.~S.~Schwarz,
  ``$A_\infty$ algebras and the cohomology of moduli spaces,''
  hep-th/9408064.

\bibitem{Gaberdiel:1997ia}
  M.~R.~Gaberdiel and B.~Zwiebach,
  ``Tensor constructions of open string theories. 1: Foundations,''
  Nucl. Phys. {\bf B505}, 569 (1997)
  [hep-th/9705038].

\bibitem{Zwiebach:1992ie}
B.~Zwiebach,
``Closed string field theory: Quantum action and the Batalin-Vilkovisky master equation,''
Nucl. Phys. B \textbf{390}, 33-152 (1993)
[arXiv:hep-th/9206084 [hep-th]].

\bibitem{Markl:1997bj}
M.~Markl,
``Loop homotopy algebras in closed string field theory,''
Commun. Math. Phys. \textbf{221}, 367-384 (2001)
[arXiv:hep-th/9711045 [hep-th]].

\bibitem{Kajiura:2003ax}
H.~Kajiura,
``Noncommutative homotopy algebras associated with open strings,''
Rev. Math. Phys. \textbf{19}, 1-99 (2007)
[arXiv:math/0306332 [math.QA]].

\bibitem{Erler:2020beb}
T.~Erler and H.~Matsunaga,
``Mapping between Witten and lightcone string field theories,''
JHEP \textbf{11}, 208 (2021)
[arXiv:2012.09521 [hep-th]].

\bibitem{Sen:2016qap}
A.~Sen,
``Wilsonian Effective Action of Superstring Theory,''
JHEP \textbf{01}, 108 (2017)
[arXiv:1609.00459 [hep-th]].

\bibitem{Erbin:2020eyc}
H.~Erbin, C.~Maccaferri, M.~Schnabl and J.~Vo\v{s}mera,
``Classical algebraic structures in string theory effective actions,''
JHEP \textbf{11}, 123 (2020)
[arXiv:2006.16270 [hep-th]].

\bibitem{Koyama:2020qfb}
D.~Koyama, Y.~Okawa and N.~Suzuki,
``Gauge-invariant operators of open bosonic string field theory in the low-energy limit,''
[arXiv:2006.16710 [hep-th]].

\bibitem{Arvanitakis:2020rrk}
A.~S.~Arvanitakis, O.~Hohm, C.~Hull and V.~Lekeu,
``Homotopy Transfer and Effective Field Theory I: Tree-level,''
Fortsch. Phys. \textbf{70}, no.2-3, 2200003 (2022)
[arXiv:2007.07942 [hep-th]].

\bibitem{Arvanitakis:2021ecw}
A.~S.~Arvanitakis, O.~Hohm, C.~Hull and V.~Lekeu,
``Homotopy Transfer and Effective Field Theory II: Strings and Double Field Theory,''
Fortsch. Phys. \textbf{70}, no.2-3, 2200004 (2022)
[arXiv:2106.08343 [hep-th]].

\bibitem{Hohm:2017pnh}
O.~Hohm and B.~Zwiebach,
``$L_{\infty}$ Algebras and Field Theory,''
Fortsch. Phys. \textbf{65}, no.3-4, 1700014 (2017)
[arXiv:1701.08824 [hep-th]].

\bibitem{Jurco:2018sby}
B.~Jur\v{c}o, L.~Raspollini, C.~S\"amann and M.~Wolf,
``$L_\infty$-Algebras of Classical Field Theories and the Batalin-Vilkovisky Formalism,''
Fortsch. Phys. \textbf{67}, no.7, 1900025 (2019)
[arXiv:1809.09899 [hep-th]].

\bibitem{Nutzi:2018vkl}
A.~N\"utzi and M.~Reiterer,
``Amplitudes in YM and GR as a Minimal Model and Recursive Characterization,''
Commun. Math. Phys. \textbf{392}, no.2, 427-482 (2022)
[arXiv:1812.06454 [math-ph]].

\bibitem{Arvanitakis:2019ald}
A.~S.~Arvanitakis,
``The $L_\infty$-algebra of the S-matrix,''
JHEP \textbf{07}, 115 (2019)
[arXiv:1903.05643 [hep-th]].

\bibitem{Macrelli:2019afx}
T.~Macrelli, C.~S\"amann and M.~Wolf,
``Scattering amplitude recursion relations in Batalin-Vilkovisky\textendash{}quantizable theories,''
Phys. Rev. D \textbf{100}, no.4, 045017 (2019)
[arXiv:1903.05713 [hep-th]].

\bibitem{Jurco:2019yfd}
B.~Jur\v{c}o, T.~Macrelli, C.~S\"amann and M.~Wolf,
``Loop Amplitudes and Quantum Homotopy Algebras,''
JHEP \textbf{07}, 003 (2020)
[arXiv:1912.06695 [hep-th]].

\bibitem{Saemann:2020oyz}
C.~Saemann and E.~Sfinarolakis,
``Symmetry Factors of Feynman Diagrams and the Homological Perturbation Lemma,''
JHEP \textbf{12}, 088 (2020)
[arXiv:2009.12616 [hep-th]].

\bibitem{Doubek:2017naz}
M.~Doubek, B.~Jur\v{c}o and J.~Pulmann,
``Quantum $L_\infty$ Algebras and the Homological Perturbation Lemma,''
Commun. Math. Phys. \textbf{367}, no.1, 215-240 (2019)
[arXiv:1712.02696 [math-ph]].

\bibitem{Masuda:2020tfa}
T.~Masuda and H.~Matsunaga,
``Perturbative path-integral of string fields and the $A_\infty$ structure of the BV master equation,''
PTEP \textbf{2022}, no.11, 113B04 (2022)
[arXiv:2003.05021 [hep-th]].

\bibitem{Batalin:1981jr}
I.~A.~Batalin and G.~A.~Vilkovisky,
``Gauge Algebra and Quantization,''
Phys. Lett. B \textbf{102}, 27-31 (1981).

\bibitem{Batalin:1983ggl}
I.~A.~Batalin and G.~A.~Vilkovisky,
``Quantization of Gauge Theories with Linearly Dependent Generators,''
Phys. Rev. D \textbf{28}, 2567-2582 (1983)
[erratum: Phys. Rev. D \textbf{30}, 508 (1984)].

\bibitem{Schwarz:1992nx}
A.~S.~Schwarz,
``Geometry of Batalin-Vilkovisky quantization,''
Commun. Math. Phys. \textbf{155}, 249-260 (1993)
[arXiv:hep-th/9205088 [hep-th]].

\bibitem{Gwilliam:2012jg}
O.~Gwilliam and T.~Johnson-Freyd,
``How to derive Feynman diagrams for finite-dimensional integrals directly from the BV formalism,''
[arXiv:1202.1554 [math-ph]].

\bibitem{Chiaffrino:2021pob}
C.~Chiaffrino, O.~Hohm and A.~F.~Pinto,
``Homological quantum mechanics,''
JHEP \textbf{02}, 137 (2024)
[arXiv:2112.11495 [hep-th]].

\bibitem{Erler:2015uba}
T.~Erler,
``Relating Berkovits and A$_\infty$ superstring field theories; small Hilbert space perspective,''
JHEP \textbf{10}, 157 (2015)
[arXiv:1505.02069 [hep-th]].

\bibitem{Okawa-Shibuya}
Y.~Okawa and S.~Shibuya,
{\it in preparation}.

\bibitem{Srednicki:2007qs}
M.~Srednicki,
``Quantum Field Theory,''
Cambridge University Press (2007).

\bibitem{Chiaffrino:2021uyd}
C.~Chiaffrino and I.~Sachs,
``QFT with stubs,''
JHEP \textbf{06}, 120 (2022)
[arXiv:2108.04312 [hep-th]].

\bibitem{Witten:2010cx}
E.~Witten,
``Analytic Continuation Of Chern-Simons Theory,''
AMS/IP Stud. Adv. Math. \textbf{50}, 347-446 (2011)
[arXiv:1001.2933 [hep-th]].

\bibitem{Okawa:2020llq}
Y.~Okawa,
``Nonperturbative definition of closed string theory via open string field theory,''
based on the talk at
{\it International Conference on String Field Theory and String Perturbation Theory}
held at The Galileo Galilei Institute for Theoretical Physics (Florence, May of 2019)
[arXiv:2006.16449 [hep-th]].

\end{thebibliography}
\end{document}